\documentclass[aps,prl,letter,twocolumn,showpacs,superscriptaddress,amsmath,amssymb,amsfonts,floatfix,citeautoscript,longbibliography]{revtex4}

\pdfoutput=1
\usepackage[T1]{fontenc}\usepackage[latin1]{inputenc}
\usepackage{dcolumn,graphicx,color,booktabs,microtype,afterpage} 
\usepackage{sidecap}
\usepackage{times}
\usepackage[colorlinks,plainpages=false,linkcolor=blue,urlcolor=blue,citecolor=blue,pdfpagemode=UseNone,pdfstartview=FitBH]{hyperref}

\usepackage{amssymb}
\usepackage{amsmath}
\usepackage{graphicx}
\usepackage{bm}
\usepackage{color}
\usepackage{stackrel}
\usepackage{enumerate}

\newcommand{\ESO}{Er$_2$Sn$_2$O$_7$}

\newcommand{\vect}[1]{\bm{#1}}

\usepackage{stackengine}
\stackMath

\begin{document}
\preprint{APS/123-QED}
%
%
\title
{Understanding Reentrance in Frustrated Magnets: the Case of the \ESO\ Pyrochlore}
\author{D. R. Yahne}
\affiliation{Department of Physics, Colorado State University, 200 W. Lake St., Fort Collins, CO 80523-1875, USA}
\author{D. Pereira}
\affiliation{Department of Physics and Astronomy, University of Waterloo, Waterloo, Ontario N2L 3G1, Canada}
\author{L. D. C. Jaubert}
\affiliation{CNRS, Universit\'{e} de Bordeaux, LOMA, UMR 5798, 33400 Talence, France}
\author{L. D. Sanjeewa}
\affiliation{Department of Materials Science and Engineering, University of Tennessee, Knoxville, TN 37996, USA}
\author{M. Powell}
\affiliation{Department of Chemistry, Clemson University, Clemson, South Carolina 29634-0973, USA}
\author{J. W. Kolis}
\affiliation{Department of Chemistry, Clemson University, Clemson, South Carolina 29634-0973, USA}
\author{Guangyong Xu}
\affiliation{NIST Center for Neutron Research, National Institutue of Standards and Technology, Gaithersburg, Maryland 20899, USA}
\author{M. Enjalran}
\affiliation{Physics Department, Southern Connecticut State University, 501 Crescent Street, New Haven, Connecticut 06515-1355, USA}
\author{M. J. P. Gingras}
\affiliation{Department of Physics and Astronomy, University of Waterloo, Waterloo, Ontario N2L 3G1, Canada}
\affiliation{CIFAR, MaRS Centre, West Tower 661 University Ave., Suite 505, Toronto, ON, M5G 1M1, Canada}
\date{\today}
\author{K. A. Ross}
\affiliation{Department of Physics, Colorado State University, 200 W. Lake St., Fort Collins, CO 80523-1875, USA}
\affiliation{CIFAR, MaRS Centre, West Tower 661 University Ave., Suite 505, Toronto, ON, M5G 1M1, Canada}
\date{\today}
%
%
%
%
\begin{abstract}
Reentrance, the return of a system from an ordered phase to a previously encountered less-ordered one as a controlled parameter is continuously varied, is a recurring theme found in disparate physical systems, from condensed matter to black holes.  While diverse in its many incarnations and generally unsuspected, the cause of reentrance at the microscopic level is often not investigated thoroughly. Here, through detailed characterization and theoretical modeling, we uncover the microscopic mechanism behind reentrance in the strongly frustrated pyrochlore antiferromagnet \ESO. Taking advantage of the recent advance in rare earth stannate single crystal synthesis, we use heat capacity measurements to expose that \ESO\ exhibits multiple instances of reentrance in its magnetic field $B$ vs. temperature $T$ phase diagram for magnetic fields along three cubic high symmetry  directions. Through classical Monte Carlo simulations, mean field theory and classical linear spin-wave expansions, we argue that the origins of the multiple occurrences of reentrance observed in  \ESO\ are linked to soft modes. Depending on the field direction, these arise either from a direct $T=0$ competition between the field-evolved ground states, or from a field-induced enhancement of the competition with a distinct zero-field antiferromagnetic phase.  In both scenarios, the phase competition enhances thermal fluctuations which  entropically stabilize a specific ordered phase. This results in an increased transition temperature for certain field values and thus the reentrant behavior. Our work represents a detailed examination into the mechanisms responsible for reentrance in a frustrated magnet and may serve as a template for the interpretation of reentrant phenomena in other physical systems.

\end{abstract}
\maketitle


Within the field of magnetism, frustration refers to a system's inability to simultaneously satisfy all of its energetic preferences. Strong frustration can result in a variety of exotic phenomena such as spin liquids, spin ice, emergent quasi-particles, topological phases and order-by-disorder~\cite{BalentsNatureQSL,gingras14a,QSLReview,Hermanns18a,HallasXYAnnuRevCMP,RauGingrasAnnuRevCMP,Knolle18a}. Most of the research focus in this area over the past thirty years has been devoted to investigating the physics near zero temperature, considering finite temperatures as a necessary {\it modus operandi} to search for signatures of the low-energy properties. However, even when subject to high frustration, a majority of frustrated magnetic materials ultimately develop long-range order or display spin-glass freezing at a nonzero critical temperature $T_\textrm{c}$, albeit often at a very low one compared to the spin-spin interactions. In this context, it therefore seems natural to ask what behavior near $T_\textrm{c}$ may be a witness of the zero-temperature ground state physics. This is particularly important when $T_\textrm{c}$ is just above the experimental baseline temperature, so that  temperatures which are low relative to $T_\textrm{c}$ cannot be reached. Here we precisely consider such a situation, as arises in the Er$_2$Sn$_2$O$_7$ pyrochlore antiferromagnet, and which provides an opportunity to study a recurrent aspect of frustrated magnetic systems observed at nonzero temperature: reentrance~\cite{Gvozdikova_2011, MnTriangle, YTOReentrance, YTOOrientation, GTOReentrance,ShannonReentrantTLFAM,Seabra16,ReentrantKitaev,Quilliam_GGG}. 

Reentrance occurs when a system, after having developed an ordered phase of some sort, returns to its original less-ordered (e.g. paramagnetic) state as some parameter (e.g. temperature, field, pressure, stoichiometry) is continuously varied. Reentrance has been found in spin glasses~\cite{ReentrantSpinGlass, Diep}, liquid mixtures~\cite{ReentrantLiquidMixtures,Vause82}, protein thermodynamics~\cite{ReentrantProtein}, liquid crystals~\cite{ReentrantLiquidCrystals,Berker81}, bilayer graphene~\cite{ReentrantBG}, superconductors~\cite{ReentrantSuperconductor}, modulated phases~\cite{mendoza19,mendoza20} and even in black hole thermodynamics~\cite{ReentrantBlackHole}.  Despite its ubiquity, reentrance is typically unexpected and its explanation in terms of entropic contributions to the free-energy from the underlying  microscopic degrees of freedom is usually subtle.  In this context, while frequently observed in frustrated magnets, the microscopic mechanism leading to reentrance often remains obscure \cite{Gvozdikova_2011, MnTriangle, YTOReentrance, YTOOrientation, GTOReentrance, ShannonReentrantTLFAM, Seabra16, ReentrantKitaev}. Two mechanisms have commonly been invoked: a field-dependent suppression of quantum fluctuations~\cite{Schmidt17,Skoulatos19,Ranjith19} and the partial disorder of an intervening phase~\cite{Vaks66,Azaria87,Boubcheur98,Diep}. Here, we present an alternative scenario of a generic nature which illustrates how the observation of reentrance may be used as a fingerprint of the frustration at play in the ground state.

\begin{figure}[ht!]
\includegraphics[scale = 1.0]{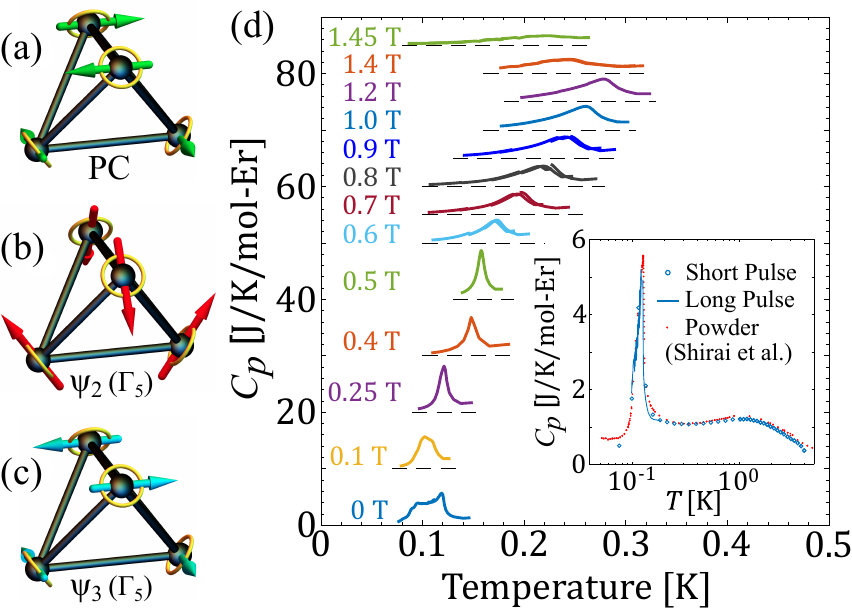}
\caption{Example of 6-fold degenerate states: (a) Palmer-Chalker~\cite{PalmerChalker} and (b) $\psi_{2}$ and (c) $\psi_{3}$ basis states of $\Gamma_5$~\cite{Poole}. The $\psi_{2}$ and $\psi_{3}$ states are connected by a rotation of the spins by an angle $\phi$ within their local easy-planes (yellow circles): $\phi \equiv n \pi/3\;(+\pi/6)$ for $n=0,...,5$ correspond to $\psi_{2}$ ($\psi_{3}$)~\cite{Savary12}. Panels (b) and (c) are for $\phi=0$ and $\pi/2$, respectively. The manifold with U(1) degeneracy, $\phi\in [0,2\pi]$, forms the so-called $\Gamma_{5}$ states that appear in the $[111]$ phase diagram. (d) Heat capacity, $C_p(T)$, vs temperature, $T$, of \ESO \ with the magnetic field along $[100]$, showing the reentrant nature of the transition.  Curves at different fields are offset vertically for clarity. (Inset) $C_p(T)$  \ESO \ in zero-field, with short and long pulse measurements on crystal samples overlaid. Powder data from Shirai {\it et al.}~\cite{Shirai2017} is also overlaid to demonstrate agreement between sample types. \label{fig_specificheat} }
\end{figure}

In this article, we show that \ESO\ represents a tractable material example where the intricate microscopic mechanisms responsible for reentrance in frustrated magnets can be rigorously studied experimentally and theoretically.  \ESO\ holds a special place among rare earth pyrochlores~\cite{HallasXYAnnuRevCMP,GardnerRMP}: it is well-characterized, has a suppressed critical temperature and is one of the few materials with a simple Palmer-Chalker (PC) antiferromagnetic ground state \cite{PalmerChalker, ESOPetitPRB, ESOPetitPRL}  [Fig.~\ref{fig_specificheat}(a)]. Its estimated exchange and single-ion susceptibility parameters are highly anisotropic and theory suggests a proximity to another competing antiferromagnetic phase~\cite{ESOPetitPRB, ESOPetitPRL,MPCPyrochlore} 
known as $\Gamma_5$~\cite{HallasXYAnnuRevCMP,RauGingrasAnnuRevCMP,MPCPyrochlore} [Fig.~\ref{fig_specificheat}(b,c)]. Because of this anisotropy, the response of \ESO\ to an applied field is expected to differ with field direction, as has been poignantly illustrated with the experimental exploration of rare earth pyrochlore titanates~\cite{GTOReentrance, YTOOrientation,HallasXYAnnuRevCMP,RauGingrasAnnuRevCMP,GardnerRMP}. As such, the present study critically relies on the recently gained ability to synthesize pyrochlore stannate single crystals~\cite{HydrothermalTin}, including \ESO{}.

\begin{figure*}[ht]
\centering\includegraphics[width=17.5cm]{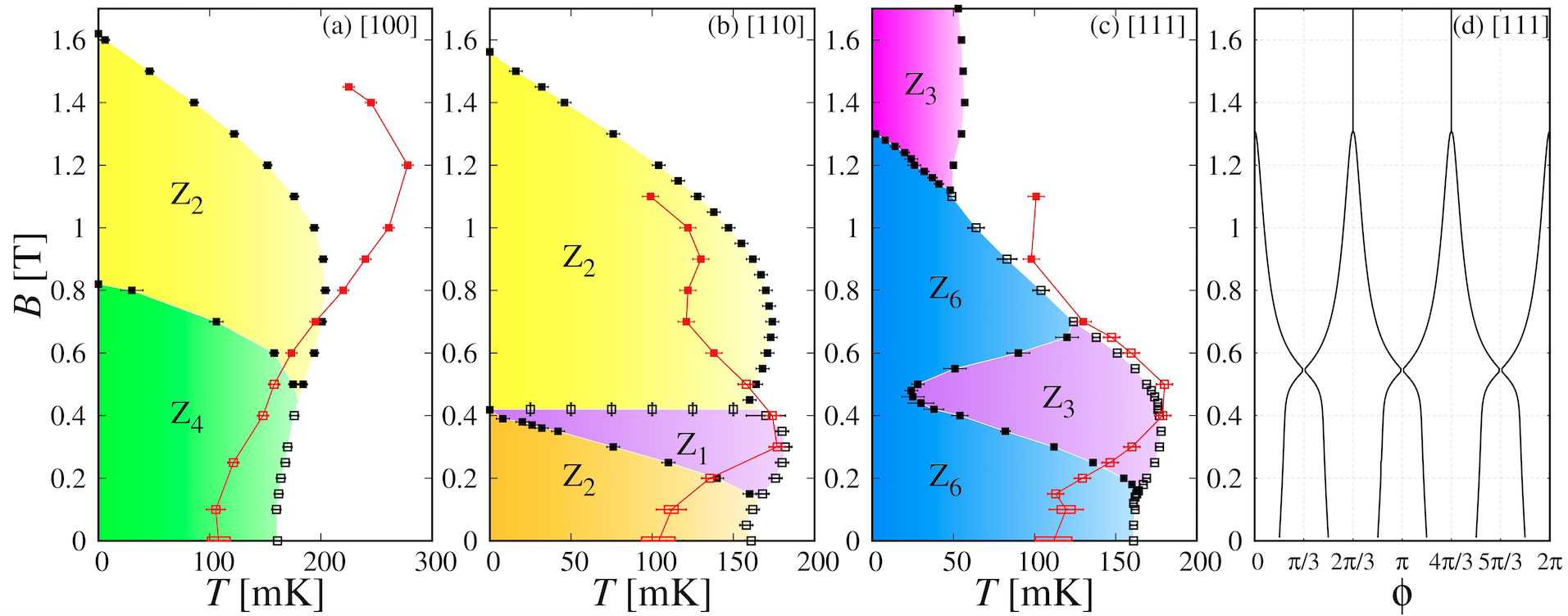}
\caption{$B$-$T$ phase diagrams of Er$_{2}$Sn$_{2}$O$_{7}$ in the (a) $[100]$, (b) $[110]$ and (c) $[111]$ field directions, comparing experimental data with sharp (\textcolor{red}{$\pmb{\square}$}) and smooth (\textcolor{red}{$\blacksquare$}) heat-capacity peaks [Fig.~\ref{fig_specificheat}(d)], to Monte Carlo results with 1$^{\rm st}$ ($\pmb{\square}$) and 2$^{\rm nd}$ ($\blacksquare$) order transitions. Experiments and simulations are notably similar, showing the same (multiple) reentrance. The degeneracy Z$_{n}$ found in simulations is given for each phase. The width of the red rectangles at $0$ and $0.1$ T represent the position of the double peaks. (d) In a $[111]$ field, each of the six FEPC ground states has a $\Gamma_{5}$ contribution described by an angle $\phi$ [Fig.~\ref{fig_specificheat}(b,c)], that can be computed exactly by minimizing the energy of one tetrahedron as a function of $B$.}
\label{fig_phasediagram}
\end{figure*}

We report herein the discovery of multiple occurrences of reentrance in the $B$-$T$ phase diagram of \ESO\ for fields along the $[100]$, $[110]$ and $[111]$ cubic directions using heat capacity measurements. By thoroughly investigating this experimental phase diagram using mean field theory, classical linear spin-wave expansions and Monte Carlo simulations, we have uncovered the various microscopic origins of reentrance in this system. In short, we find that different types of multi-phase competitions at $T=0$ result in enhanced thermal fluctuations that \textit{entropically} stabilize the \textit{ordered} phase, and thus increase $T_\textrm{c}(B)$ over a certain $B$ field range.  This produces $T_\textrm{c}(B)$ reentrant phase boundaries whose maximal temperature extent corresponds to the zero-temperature field-driven phase transitions [see Fig.~\ref{fig_phasediagram}(a-c)]. This multi-phase competition is in some cases a direct consequence of the competition of the field-evolved  PC states while in others it is inherent to \ESO's zero-field ground state being  in close proximity to the phase boundary between the PC and $\Gamma_5$ phases. See the Supplementary Material~\cite{SuppMat}
for technical details on the experiments, simulations and analytics.\\


 Heat capacity ($C_p$) measurements were performed on single crystal samples, grown via the hydrothermal method described in 
Ref.~\cite{HydrothermalTin}, down to $50$ mK with varying magnetic fields, ${\bm B}$, oriented in the $[111]$, $[110]$, and $[100]$ directions, using a dilution refrigerator insert in a Quantum Design Physical Properties Measurement System. Two measurement techniques were used: the conventional quasi-adiabatic thermal relaxation method (called "short pulses" hereafter), as well as "long pulses", both of which are described in detail in Ref.~\cite{Scheie2018}. The long pulse technique allows faster and higher point-density measurements across phase transitions, enabling an accurate  mapping of a phase diagram by measuring the field dependence of the transition temperature, $T_\textrm{c}(B)$ [Fig.~\ref{fig_specificheat} (d)].

In the zero-field heat capacity [Fig.~\ref{fig_specificheat} (d) inset], we find a sharp lambda-like anomaly indicative of a phase transition at $T_\textrm{c} = 118 \pm 5$ mK, which is consistent with previous  measurements on powder samples reported in Ref.~\cite{Shirai2017} (130 mK, from heat capacity, data shown in Fig.~\ref{fig_specificheat} (d) inset) and Ref.~\cite{ESOPetitPRL} ($108 \pm 5$ mK, from DC magnetic susceptibility). The extremely high point density of the long pulse measurements allow for the observation of subtle features in the peak shape, which are typically not resolved by conventional short pulse measurements.  This reveals a low temperature shoulder of the $C_p$ peak in the zero-field data at $97 \pm 5$ mK \cite{SuppMat}. We performed elastic neutron scattering measurements to determine the magnetic structure between the sharp high temperature peak and the low temperature shoulder to check for an intermediate magnetic phase~\cite{SuppMat}.  We found that the magnetic structure  is of Palmer-Chalker type  at all measured temperatures throughout the transition range with no sign of other magnetic phases. It is not clear what causes this structure in the heat capacity anomaly, but we note that similar (though not identical) broadening is observed in all five crystals we have measured as well as in published data on a powder sample~\cite{Shirai2017, SuppMat} (Fig.~\ref{fig_specificheat} (d) inset).  Thus, it seems to be a feature of all \ESO \ samples, but is likely due to (or influenced by) slight inhomogeneities rather than being purely intrinsic in origin.  Although it may be worth future investigation, its presence does not affect any of the conclusions of this work.\\


To model Er$_{2}$Sn$_{2}$O$_{7}$, we use the generic nearest-neighbor Hamiltonian on the pyrochlore lattice \cite{Ross2011,MPCPyrochlore},
\begin{eqnarray}
\mathcal{H} & = & \sum_{\langle i,j\rangle} 
{J}_{ij}^{\alpha\beta}
{S}_{i}^\alpha  
{S}_{j}^\beta   
\;-\; \mu_{\rm B} \, \sum_{i} \, 
{{g}}_{i}^{\alpha\beta} 
{B^\alpha} 
{S}_i^\beta.
\label{eq:ham}
\end{eqnarray}
${\bm S}_i = (S^x_i, S^y_i, S^z_i)$ is a three-component pseudo-spin of length $|{\bm S}_{i}|=1/2$ and ${\bm B}$ is the external magnetic field. The $g$-tensor represents the single-ion anisotropy, with local easy-plane $g_{\perp}$ and easy-axis $g_{\parallel}$ components at lattice site $i$. Given the symmetries of the pyrochlore lattice, the anisotropic exchange matrix $J_{ij}^{\alpha \beta}$ is parameterized by four independent coupling constants: $(J_{1}, J_{2}, J_{3}, J_{4})$~\cite{Curnoe2007,Ross2011}. Er$_{2}$Sn$_{2}$O$_{7}$ has been previously parameterized using inelastic neutron scattering on powder samples~\cite{ESOPetitPRB,ESOPetitPRL}. Here we choose to remain within the error bars of Ref.~\cite{ESOPetitPRL}, selecting a set of coupling parameters where simulations find $T_\textrm{c}\sim 180$ mK at $0.4$ T for a $[111]$ field to match the experimental result~\footnote{Now that single crystals are available, fine-tuning these parameters by exploring the entire $Q-$space with neutron scattering would be a worthwhile endeavor for future work, albeit challenging due to the small mass of each crystal.}: $(J_{1}, J_{2}, J_{3}, J_{4})=(+0.079, +0.066, -0.111,+0.032)$ meV and $g_{\perp}=7.52, g_{\parallel}=0.054$.

To proceed, we first analyze this model using classical Monte Carlo simulations, with the results  summarized in the $B$-$T$ phase diagrams of Fig.~\ref{fig_phasediagram}.  Most importantly, with $T_\textrm{c}(B\!=\!0.4\,{\rm T})$ fitted (for $\bm{B}$ along the [111] direction),  the simulations reproduce the number of reentrant ``lobes'' for each field direction (e.g. one and two for a $[100]$ and $[110]$ field, respectively), as well as, at each lobe, the rough magnitude of the increase of $T_\textrm{c}$ at the corresponding value of $B$. Moreover, simulations find that the transition always evolves from discontinuous to continuous when increasing the field. This is consistent with the shape of the experimental heat capacity peaks, evolving from sharp to smooth [Fig.~\ref{fig_specificheat}(d) and Ref.~\cite{SuppMat}]. We suspect that a fine-tuning of the ${J}_{ij}^{\alpha\beta}$ coupling parameters and  incorporating quantum fluctuations and dipolar interactions should account for the quantitative disagreements. Nevertheless, the  semi-quantitative match between experiments and simulations confirms the validity of Eq.~\eqref{eq:ham} as a minimal model for \ESO, suggesting that it robustly encapsulates the key physics behind the experimentally observed multiple occurrences of reentrance.

The results in Figs.~\ref{fig_phasediagram}(a-c) raise multiple questions.  Why are there  multiple instances of reentrance and why are they  so strongly dependent on the field direction?  More fundamentally, why does \ESO\ demonstrate reentrance in the first place? As a set of clues, simulations bring to light a variety of phases that vie for ordering. In the rest of this article, we explain how soft modes induced by this multi-phase competition are linked to reentrance, using a combination of mean field theory and classical linear spin-wave expansions.

The zero-field ground state of \ESO\ is the sixfold-degenerate PC phase. However, the ground states naturally deform and evolve under the application of a magnetic field. For sufficiently large fields, some of these field-evolved PC (FEPC) states may become partially polarized into the \textit{same} spin configuration. We therefore label the resulting phase according to the number of FEPC states that minimize the free energy but have \textit{distinct} spin configurations (e.g. Z$_6$ at $B = T = 0$, for the six degenerate PC states, and Z$_1$ at sufficiently large $B$ for the trivial field-polarized paramagnet). Phase transitions then occur whenever distinct FEPC states ``merge'' into the same spin configuration at a given field value $B_\textrm{c}$.

%

\begin{figure}[t]
\centering\includegraphics[width=8.75cm]{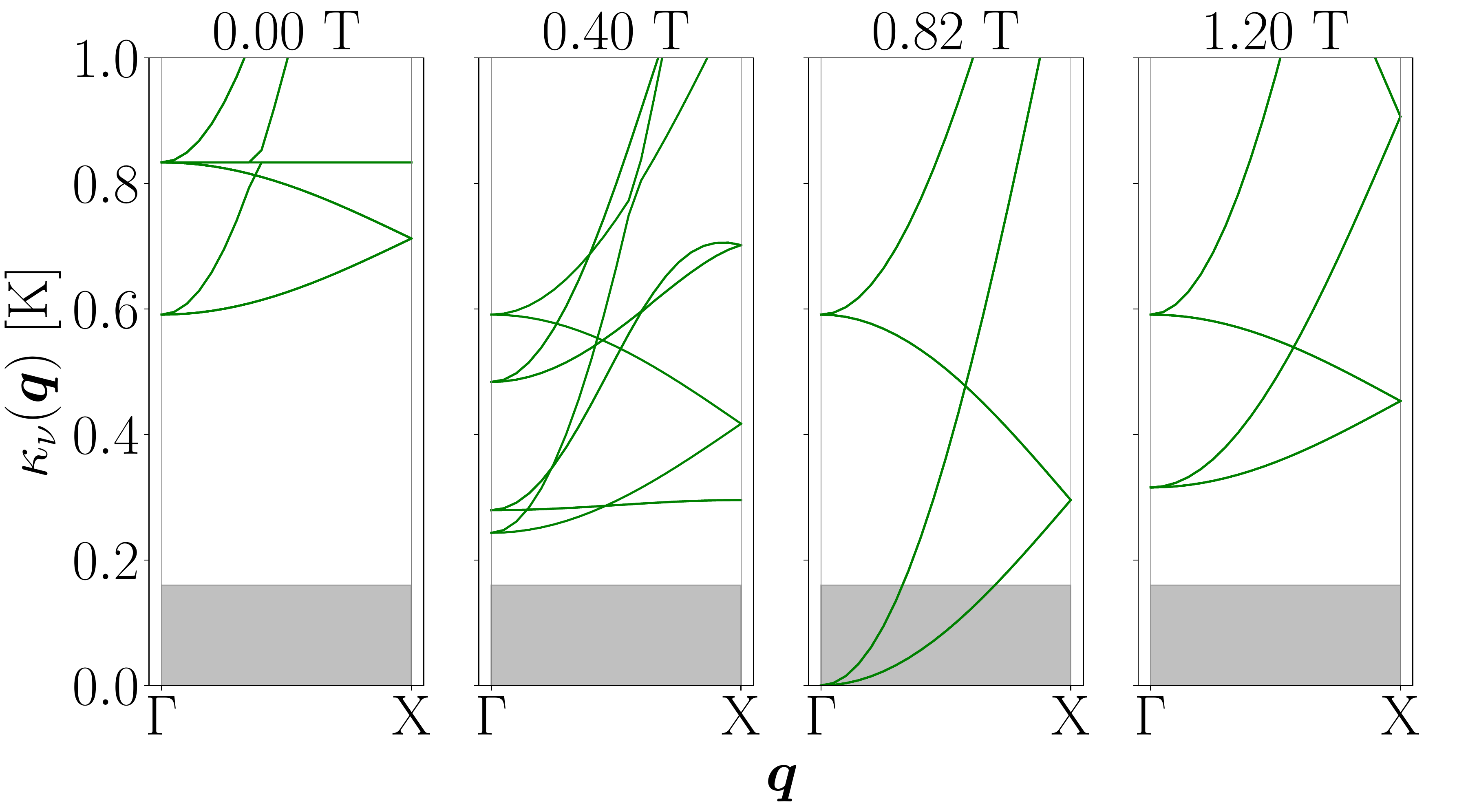}
\caption{Classical spin-wave dispersions for $B = 0, 0.40, 0.82$, and $1.20$~T along the $[100]$ direction, for a path in the FCC Brillouin zone. Note that $B_\textrm{c} = 0.82$~T is a critical field at $T=0$, as shown in Fig. \ref{fig_phasediagram}. The grey boxes indicate energy scales below $T_\textrm{c}(B = 0\ {\rm T}) \approx 160$~mK from Monte Carlo simulations; when modes occur within this region they are considered ``soft''. Note that the dispersions for \textit{all} Palmer-Chalker states are plotted, but may overlap at high-symmetry points or due to their degeneracies in a field.}
\label{fig:SWT}
\end{figure}

First, consider the $[100]$ field phase diagram in Fig.~\ref{fig_phasediagram}(a).  
At $T=0$, the FEPC states merge at $B_\textrm{c} = 0.82$ T, giving rise to the yellow Z$_{2}$ region. Fig. \ref{fig:SWT} displays the classical spin-wave dispersions $\kappa_\nu(\vect{q})$ for a number of field values below and above $B_\textrm{c}$, calculated from the corresponding $T=0$ FEPC ground states. As the merger transition is approached at $B_{{\textrm c}} = 0.82$ T, the bottom of the dispersive bands drop  below the energy scale set by $T_\textrm{c}(B = 0) \approx 160$ mK, becoming soft and gapless at $B = B_\textrm{c}$. This decrease indicates a propensity for stronger thermal fluctuations at $B_\textrm{c}$ than at other field values. More precisely, since $s = -\frac{1}{8N_q}\sum_{\vect{q}} \sum_{\nu=1}^8 \ln(\kappa_{\nu}(\vect{q}))$ 
quantifies the entropy contribution from classical spin-waves, the decrease in $\kappa_\nu(\vect{q})$ on approaching $B_\textrm{c}$ from above or below (as shown in Fig. \ref{fig:SWT}) corresponds to an \textit{increase} in entropy within the ordered phase. As a consequence, the gapless soft modes at $B_\textrm{c}$ stabilize the yellow Z$_{2}$ region of Fig.~\ref{fig_phasediagram} at finite temperature, both over the green Z$_{4}$ region as well as the disordered paramagnet. This increase in entropy of the ordered phase due to the merger transition is the reason why reentrance occurs near $B_\textrm{c}$;  the gapless soft modes resulting from this merger transition are thus responsible for the reentrance. 

 The picture is different in a $[111]$ field; simulations reveal a reentrant lobe around a field value ($\sim$ $0.4$ T) for which no corresponding $T=0$ FEPC merger is found in the calculations. To understand this reentrance, it is important to note that the long-range order of \ESO\ in a $[111]$ field is not described by a single irreducible representation (irrep) \cite{MPCPyrochlore}. Instead, it is described by the naturally field-induced ferromagnetic irrep as well as the $\Gamma_5$ irrep [Fig.~\ref{fig_specificheat}(b,c)] due to the proximity of the $\Gamma_5$ ground state to the PC phase in zero- field~\cite{MPCPyrochlore}. The $\Gamma_5$ states bear an accidental U(1) degeneracy parameterized by an angle $\phi$~\cite{Champion,Savary12,Zhitomirsky12,Oitmaa13,MPCPyrochlore}, which is lifted by a magnetic field~\cite{Maryasin16} with  discrete values of $\phi$ being selected, as shown in Fig.~\ref{fig_phasediagram}(d). While the six FEPC states remain distinct in this region, their $\Gamma_5$ components merge at $B \approx 0.55$ T into three  $\phi = \{\pi/3, \pi, 5\pi/3\}$ corresponding to $\psi_2$ states [Fig.~\ref{fig_specificheat}(b), Ref.~\cite{SuppMat}]. This $\psi_2$ selection is associated with a flat low-energy soft mode in the spin-wave expansion at $B \approx 0.5$ T and simulations confirm the presence of partial $\psi_{2}$ order in the reentrant lobe, shown by the violet Z$_3$ phase in Fig.~\ref{fig_phasediagram}(c). These results make a strong case unraveling the mechanism of the reentrance; the intervening Z$_3$ phase is entropically stabilized by low-energy soft modes arising from the PC and $\Gamma_5(\psi_2)$ phase competition.

Closing the $[111]$ case, one should mention the FEPC merger transition at $B_\textrm{c} = 1.31$ T and $T=0$ is naturally accompanied by a merging of the $\phi$ values (here also corresponding to $\psi_2$ states, see Fig.~\ref{fig_phasediagram}(d)) and by a small reentrant lobe [Fig.~\ref{fig_phasediagram}(c)], as expected from the discussion for the $[100]$ field case. Our experimental data point towards the onset of this high-field lobe as well (see Fig.~\ref{fig_phasediagram}(c) for $B= 0.9$ T \& 1.1 T). However, it was not possible to explore this high-field region experimentally because the sample did not easily equilibrate above $0.7$ T. Interestingly, simulations also suffer from difficulties thermalizing between $0.7$ and $1.2$ T.

Finally, the mechanisms behind reentrance for a $[110]$ field are reminiscent of the other two field directions [Fig.~\ref{fig_phasediagram}(b)]. Below $B\lesssim 0.1$ T, simulations are difficult to thermalize, but above $B\gtrsim 0.1$ T, we find two FEPC ground states that merge at $B_\mathrm{c}=0.42$ T. This merging gives rise to gapless soft modes, the subsequent violet Z$_1$ phase, and reentrance at finite temperatures. It is the same mechanism as in a $[100]$ field. However, as opposed to the $[100]$ scenario, this newly merged ground state vanishes immediately once $B>B_{\textrm {c}}$ (i.e. it becomes an excited state). The system is then found in two \textit{other} ground states (corresponding to the yellow Z$_2$ phase). The vanishing of the merged state corresponds to the Z$_1$ phase abruptly disappearing above $B_{\textrm{c}}$ and the removal of the aforementioned gapless soft modes. This causes the rapid collapse of the reentrant lobe at $B\sim 0.42$ T. At higher field, there are no ground state FEPC mergers, but the gap in spin-wave excitations remains small for a broad field region around $0.7$ T 
(see Section S5 in Ref~\cite{SuppMat}).
These low-energy modes, concomitantly with the collapse at $B_\textrm{c}=0.42$ T, are the reason for the higher field reentrant lobe at $\sim 0.7$ T.


In summary, we have presented the first exploration of the field-direction dependence of the thermodynamics of stannate pyrochlores, which, despite decades of effort, were not available as single crystals until very recently \cite{HydrothermalTin}. Access to these crystals has proven to be crucial since the phase diagram of \ESO\ is highly sensitive to the field direction, and exhibits several reentrant lobes with sundry underlying mechanisms. These features result from the competition of several orders, especially the zero-field Palmer-Chalker, the field-induced ferromagnetic and the neighboring $\Gamma_5$ states. In particular, most instances of reentrance in the phase diagram can be traced to zero-temperature field-induced merging of distinct degenerate states, leading to soft modes which entropically enhance the transition temperatures. In this light, reentrance is a useful and experimentally accessible fingerprint of a zero-temperature phase transition.

Given that multi-phase competition is a common feature of frustrated magnetism, we expect the mechanisms we have uncovered to be widespread among magnetic systems displaying reentrance; especially since it does not require the accidental presence of an exotic partially-disordered phase \cite{Azaria87,Diep}. In semi-classical and quantum systems, our mechanism may work together with the field-induced suppression of quantum fluctuations~\cite{Schmidt17} to produce even larger reentrant lobes. We hope our work will motivate others to pursue a microscopic interpretation of future observations of reentrance (and possibly to revisit old ones \cite{Gvozdikova_2011, MnTriangle, YTOReentrance, YTOOrientation, GTOReentrance, Quilliam_GGG, ShannonReentrantTLFAM, Seabra16,ReentrantKitaev}) in light of zero-temperature transitions. Since magnetic systems often afford us with minimal models to understand other areas of physics, our results raise a more general question: if reentrance is observed by varying a given parameter, when is it actually due to a nearby transition in a broader parameter space? \\

\begin{acknowledgements}
We acknowledge Natalia Perkins for useful discussions and Rob Mann for comments on reentrance in black holes. We thank Allen Scheie and Tom Hogan for their help with analyzing the long pulse measurements, and acknowledge the use of the LongHCPulse program for this analysis. We thank I. Zivkovic and R. Freitas for providing us their $C_p$ data previously published in Ref.\cite{Shirai2017}. We acknowledge the support of the National Institute of Standards and Technology, U.S. Department of Commerce, in providing the neutron research facilities used in this work. This research was partially supported by CIFAR.   DRY, KAR, and JWK acknowledge funding from the Department of Energy award DE-SC0020071 during the preparation of this manuscript.   The work at the University of Waterloo was supported by the Natural Sciences and Engineering Research Council (NSERC) and by the Canada Research Chairs Program (M.G, Tier 1). L. J. acknowledges financial support from CNRS (PICS No. 228338) and from the French ``Agence Nationale de la Recherche'' under Grant No. ANR-18-CE30-0011-01. 
 \end{acknowledgements}

%
%
%
%


\end{document}


%
%
%
%
\title
{Supplemental Material for: ``Understanding Reentrance in Frustrated Magnets: the Case of the \ESO\ Pyrochlore''}
\author{D. R. Yahne}
\affiliation{Department of Physics, Colorado State University, 200 W. Lake St., Fort Collins, CO 80523-1875, USA}
\author{D. Pereira}
\affiliation{Department of Physics and Astronomy, University of Waterloo, Waterloo, Ontario N2L 3G1, Canada}
\author{L. D. C. Jaubert}
\affiliation{CNRS, Universit\'{e} de Bordeaux, LOMA, UMR 5798, 33400 Talence, France}
\author{L. D. Sanjeewa}
\affiliation{Department of Materials Science and Engineering, University of Tennessee, Knoxville, TN 37996, USA}
\author{M. Powell}
\affiliation{Department of Chemistry, Clemson University, Clemson, South Carolina 29634-0973, USA}
\author{J. W. Kolis}
\affiliation{Department of Chemistry, Clemson University, Clemson, South Carolina 29634-0973, USA}
\author{Guangyong Xu}
\affiliation{NIST Center for Neutron Research, National Institutue of Standards and Technology, Gaithersburg, Maryland 20899, USA}
\author{M. Enjalran}
\affiliation{Physics Department, Southern Connecticut State University, 501 Crescent Street, New Haven, Connecticut 06515-1355, USA}
\author{M. J. P. Gingras}
\affiliation{Department of Physics and Astronomy, University of Waterloo, Waterloo, Ontario N2L 3G1, Canada}
\author{K. A. Ross}
\affiliation{Department of Physics, Colorado State University, 200 W. Lake St., Fort Collins, CO 80523-1875, USA}
\affiliation{CIFAR, MaRS Centre, West Tower 661 University Ave., Suite 505, Toronto, ON, M5G 1M1, Canada}
\date{\today}
\maketitle

\renewcommand{\thesection}{S\arabic{section}}   \renewcommand{\theequation}{S\arabic{equation}}
\renewcommand{\thefigure}{S\arabic{figure}}
\renewcommand{\thetable}{T\arabic{table}}

\numberwithin{equation}{section}
\numberwithin{figure}{section} \numberwithin{table}{section} 

This document is a supplement to our main article and contains a number of brief sections to assist the reader with
details. Specifically, we provide in turn:
\begin{itemize}
    \item Section~\ref{SM-PC}: A definition of the Palmer-Chalker states.
    \item Section~\ref{SM-AddCp}: Additional heat capacity data for the different field directions.
    \item Section~\ref{SM-Neutronexp}: Neutron scattering experimental details and results.
    \item Section~\ref{SM-TechMFT}: Details of the mean field theory used to describe the field-induced Palmer-Chalker state merger transitions.
    \item Section~\ref{SM-TechSW}: Details of the classical spin-wave expansion.
    \item Section~\ref{SM-MC}: Details of the Monte Carlo simulations and choice of model Hamiltonian parameters.
    \item Section~\ref{SM-Analysis110MC}: Details of the analysis for the phase diagram for fields along the $[110]$ direction.
       \item Section~\ref{SM-Analysis110MCsmallfield}: Details of the Monte Carlo simulations at low fields along the $[110]$ direction.
    \item Section~\ref{SM-Analysis111MC}: Details of the Monte Carlo simulations for fields along the $[111]$ direction.
\end{itemize}

\newpage

\section{Definition of Palmer-Chalker States}
\label{SM-PC}

\begin{figure}[ht!]
\includegraphics[width=12cm]{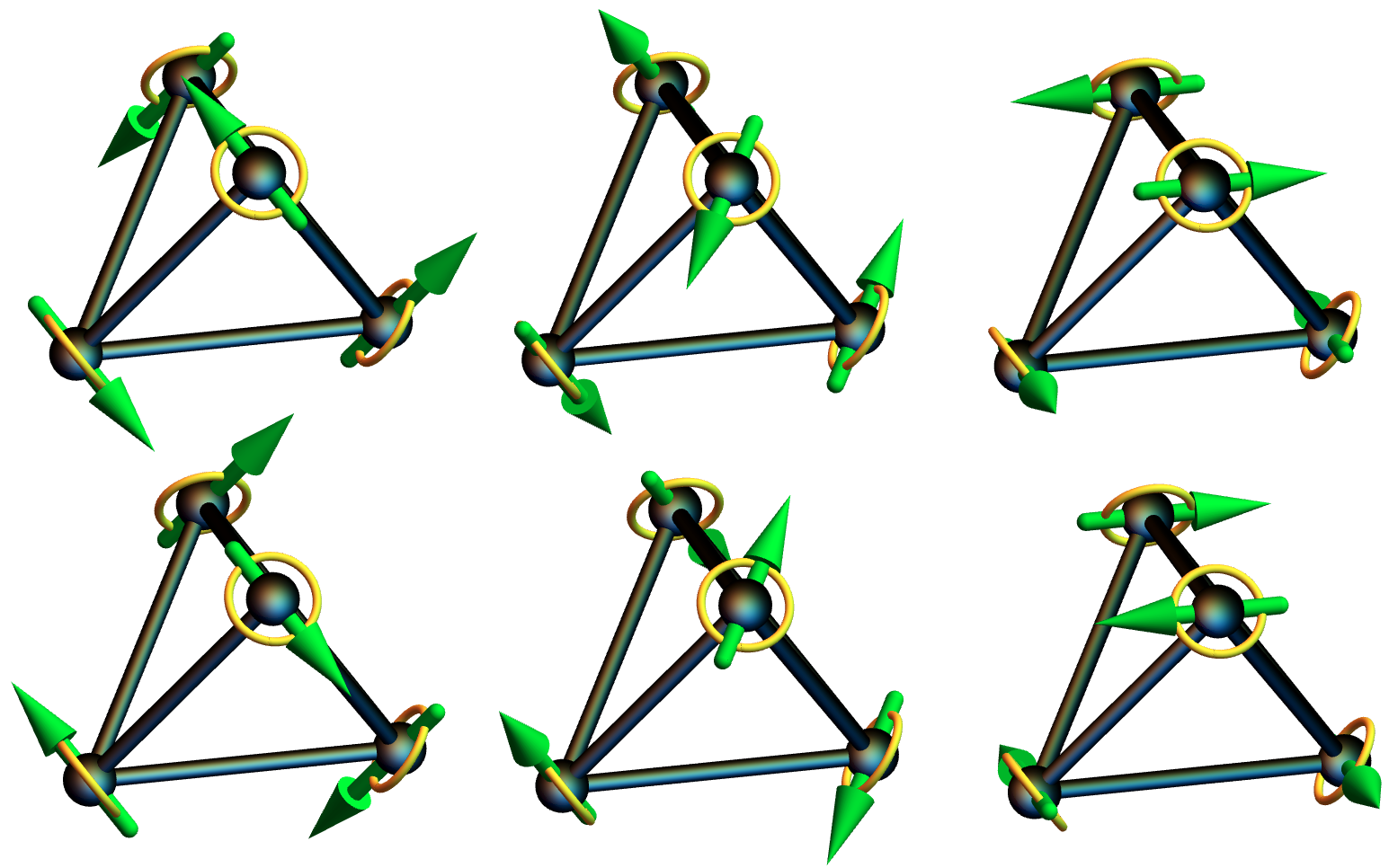}
\caption{The six Palmer-Chalker states, labeled from left to right $\langle xz \rangle$, $\langle yz \rangle$, $\langle xy \rangle$ (first row) and $\langle \overline{xz} \rangle$, $\langle \overline{yz} \rangle$, $\langle \overline{xy} \rangle$ (second row).
\label{fig_6PC} }
\end{figure}

There are six Palmer-Chalker states in zero field [Fig.~\ref{fig_6PC}]; the spin configurations for three of these (on the four sublattices of a tetrahedron) are outlined in Table \ref{tbl:PCStates}. The remaining three, denoted as $\langle \overline{xy} \rangle$, $\langle \overline{xz} \rangle$, and $\langle \overline{yz} \rangle$, can be obtained from the listed three by reversal of the spins. 

\begin{table}[h]
    \begin{tabular}{|| c | c | c ||}
         \hline $\langle xy \rangle$ & $\langle xz \rangle$ & $\langle yz \rangle$ \\
         \hline \hline
         $\vect{m}_0 = \frac{1}{\sqrt{2}}(1,-1,0)$ & $\vect{m}_0 = \frac{1}{\sqrt{2}}(1,0,-1)$ & $\vect{m}_0 = \frac{1}{\sqrt{2}}(0,1,-1)$ \\
         $\vect{m}_1 = \frac{1}{\sqrt{2}}(-1,-1,0)$ & $\vect{m}_1 = \frac{1}{\sqrt{2}}(-1,0,-1)$ & $\vect{m}_1 = \frac{1}{\sqrt{2}}(0,-1,1)$ \\
         $\vect{m}_2 = \frac{1}{\sqrt{2}}(1,1,0)$ & $\vect{m}_2 = \frac{1}{\sqrt{2}}(-1,0,1)$ & $\vect{m}_2 = \frac{1}{\sqrt{2}}(0,-1,-1)$ \\
         $\vect{m}_3 = \frac{1}{\sqrt{2}}(-1,1,0)$ & $\vect{m}_3 = \frac{1}{\sqrt{2}}(1,0,1)$ & $\vect{m}_3 = \frac{1}{\sqrt{2}}(0,1,1)$ \\
         \hline \hline
    \end{tabular}
    \caption{Spin configurations for three of the six Palmer-Chalker states on the four sublattices $i = 0, 1, 2, 3$ of a tetrahedron. Each state lies within a plane in the global frame of reference (e.g. $\langle xy \rangle$ lies in the $xy$-plane). The remaining three Palmer-Chalker states are obtained by spin reversal. The convention for labeling the sublattices follows the one from Ref. [\onlinecite{MPCPyrochlore}].}
    \label{tbl:PCStates}
\end{table}
\clearpage
\section{Additional Heat Capacity Data}
\label{SM-AddCp}

\begin{figure*}[ht!]
\includegraphics[scale = 1.5]{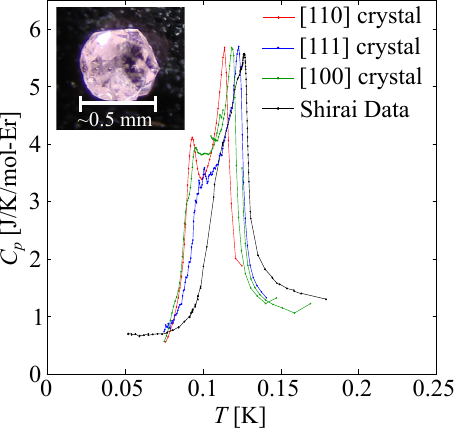}
\caption{Raw zero field heat capacity data for the three crystals polished in each of the high symmetry directions. We find that all samples show a low temperature shoulder, but the pronouncement of the shoulder varies between samples. Inset: Example of a typical crystal that was used in heat capacity measurements. \label{appfig_zerofield}}
\end{figure*}

\begin{figure*}[ht!]
\includegraphics[scale = 1.0]{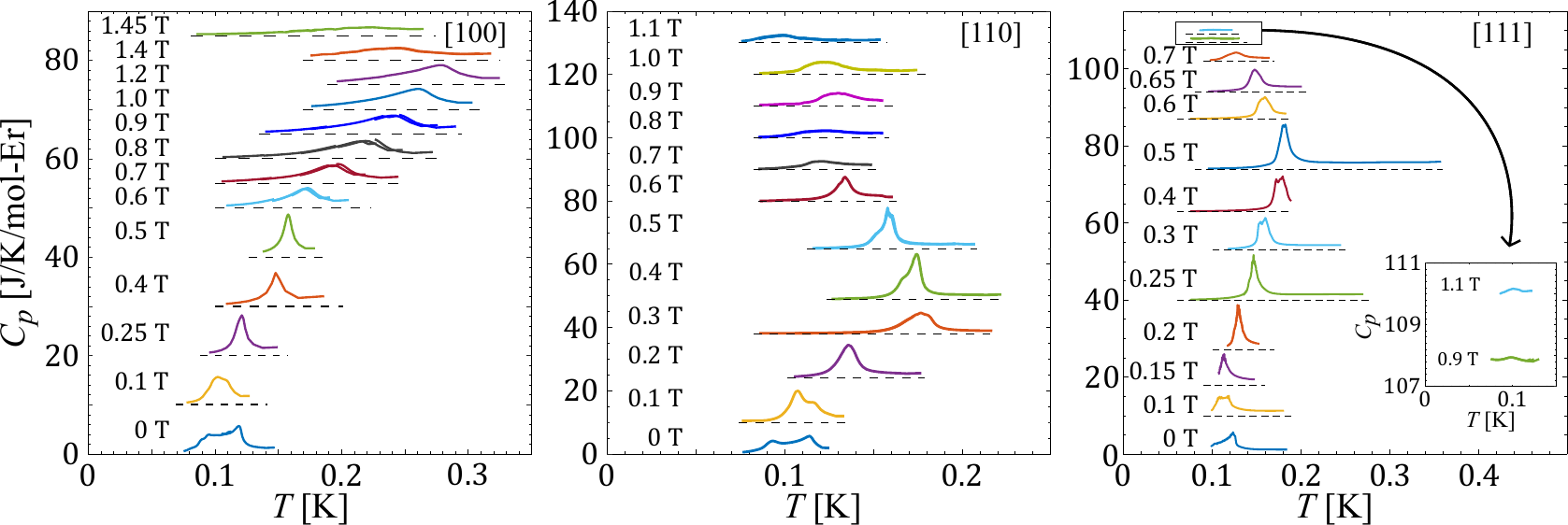}
\caption{Raw long pulse heat capacity data for fields along the three high symmetry cubic directions, leading to the phase diagram shown in the main text. Note, all data at nonzero field have been shifted vertically for clarity. \label{appfig_PhD}}
\end{figure*}

Figure~\ref{appfig_zerofield} shows the raw zero-field long pulse heat capacity data for three different single crystals polished in the three high symmetry directions. As discussed in the main text, we find a sharp peak at $118 \pm 5$ mK with a low-temperature shoulder around $97 \pm 5$ mK. A similar broad feature can be seen in polycrystalline measurements by Ref.~[\onlinecite{Shirai2017}], but the broadness of the peak is not discussed therein. The heat capacity measurements in field are shown in Fig.~\ref{appfig_PhD} for fields applied along the three cubic directions, the transition temperatures of which are included in Fig. 2 in the main text. 

\clearpage

\section{Neutron Scattering Experimental Details and Results}
\label{SM-Neutronexp}

Elastic neutron scattering measurements were performed at the NIST Center for Neutron Research using the SPINS triple axis spectrometer. Approximately $4$ g of crystals were ground into a fine powder and placed in a copper cigar type foil construction that was inside the sample can which was subsequently filled with $10$ atm of He gas at room temperature. We collected elastic scattering scans at $50$ mK, $110$ mK, and $8$ K on the $(111)$, $(002)$, $(220)$, $(113)$ and $(222)$ Bragg peaks using $E = 5$ meV neutrons with an energy resolution of approximately $0.25$ meV. The collimation settings were guide-open-$80'$-open, with Be filters placed before and after the sample.

\begin{figure*}[ht!]
\includegraphics[scale = 1.3]{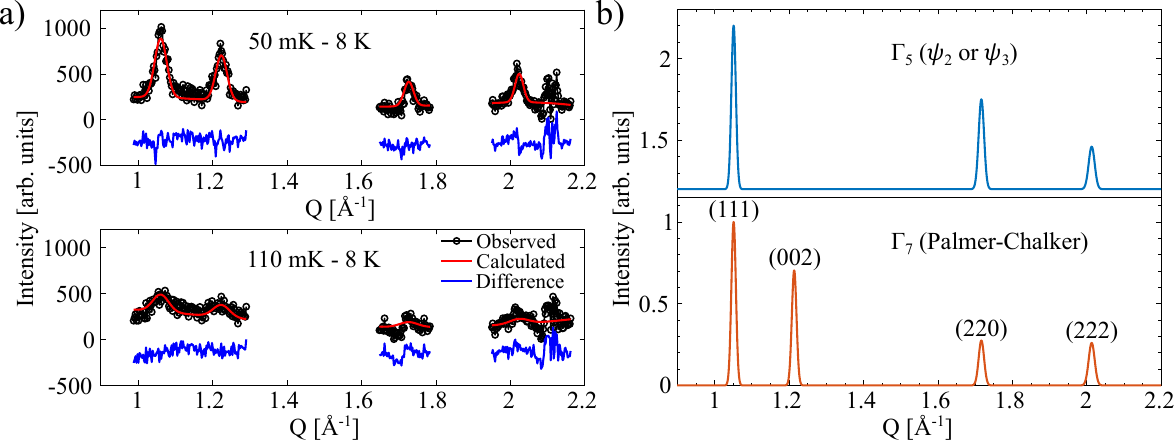}
\caption{ (a) Elastic neutron scattering data (black) on SPINS. Data was taken at $50$ mK and $110$ mK, and $8$ K data was used for background subtraction. The fit to the PC phase is shown (red), as well as the difference between the data and fit (blue). (b) Calculated magnetic diffraction patterns for the $\Gamma_5$ and PC ($\Gamma_7$)  configurations (nuclear contributions not shown). The $\Gamma_5$ pattern has been shifted vertically for clarity. \label{appfig_SPINS}}
\end{figure*}

The $50$ mK data was taken to corroborate the Palmer-Chalker ground state found from powder samples by Ref.~[\onlinecite{ESOPetitPRL}], as it should be well into the ordered phase. Data at $110$ mK was taken as it lies in between the high temperature and low temperature peaks we find in the heat capacity curve ($118$ and $97$ mK respectively), as discussed in the main text, and could possibly show evidence for an intermediate phase. We find that the correlations are Palmer-Chalker throughout the broad transition, as evidenced by the ${\textrm{Q}}=(111), (002)$ and $(220)$ Bragg peaks. In particular, there was no hint of $\Gamma_5$ order, which may have been a likely intermediate phase candidate due to the material's proposed proximity to the phase boundary between PC and $\Gamma_5$~[\onlinecite{ESOPetitPRB,ESOPetitPRL}]. No $(002)$ intensity is expected for the $\Gamma_5$ phase (Fig~\ref{appfig_SPINS} (b)), and we do not see any evidence of an enhanced $(111)$ nor $(220)$ peak that could indicate coexistence of the $\Gamma_5$ and PC ($\Gamma_7$) phase. The fit to the Palmer-Chalker structure is shown in Fig~\ref{appfig_SPINS} (a) and is in agreement with the Ref. [\onlinecite{ESOPetitPRL}] powder diffraction data. We note that only intermediate phases with ordering wave vector ${\bm q}=0$ were investigated, and it is still possible that a ${\bm q} \ne 0$  intermediate phase could exist.

\clearpage
\section{Variational Mean Field Theory and Palmer-Chalker Merger Transitions}
\label{SM-TechMFT}

The following derivation of variational mean field theory for a spin system on the pyrochlore lattice, with an exchange Hamiltonian and an applied magnetic field, follows Refs. [\onlinecite{VMFT91, VMFT03, VMFT}]. The exchange Hamiltonian can be written as
\begin{equation}
\mathcal{H}_0 = \frac{1}{2} \sum_{a,b} \sum_{i,j} \sum_{\mu,\nu} J_{ia, jb}^{\mu \nu}S_{ia}^{\mu}S_{jb}^{\nu},
\end{equation}
where $a$ and $b$ denote the chosen tetrahedra, $i$ and $j$ denote the sublattices chosen within those tetrahedra, and $\mu$ and $\nu$ denote the components of the (classical) spin vectors. The exchange matrix $J_{ia,jb}^{\mu \nu}$ is completely general in this form. In this work, we have predominantly only taken into account the global-frame nearest-neighbor exchange parameters that are defined for rare-earth pyrochlore systems with a Kramers ground state doublet [\onlinecite{MPCPyrochlore}]. However, this general form can also include the dipole-dipole interaction (see further below at the end of this section). With an applied magnetic field, this Hamiltonian becomes
\begin{equation}
\mathcal{H} = \frac{1}{2} \sum_{a,b} \sum_{i,j} \sum_{\mu,\nu} J_{ia,jb}^{\mu \nu}S_{ia}^{\mu}S_{jb}^{\nu} - \sum_{i,a} \vect{h}_{ia} \cdot \vect{S}_{ia},
\end{equation}
where
\begin{equation}
h_{ia}^{\mu} = \mu_{\textrm B} g_{ia}^{\mu \nu} B^{\nu} \label{eq:ZeemanField}
\end{equation}
is the scaled magnetic field for a given applied magnetic field $B^{\nu}$, $g_{ia}^{\mu \nu}$ is the $g$-tensor of the ion at the given sublattice $i$, expressed in the global frame of Ref. [\onlinecite{MPCPyrochlore}] and $\mu_{\textrm B}$ is the Bohr magneton.  This $g$-tensor is independent of the tetrahedron $a$ and only depends on which sublattice $i$ is considered [\onlinecite{MPCPyrochlore}].  Henceforth, all physical quantities with energy dimension are measured 
in K (Kelvin) units and, as such, we correspondingly set the Boltzmann constant $k_{\textrm B}=1$.

The mean field free energy is given by
\begin{equation}
F_{\rho} = \textrm{Tr}\{\rho \mathcal{H}\} + T\textrm{Tr}\{\rho \ln \rho \},
\end{equation}
where $\rho$ is the many-body density matrix and the trace is computed over all spin configurations. The variational mean field approximation assumes $\rho (\{\vect{S}_{ia}\}) = \prod_{i,a} \rho_{ia} (\vect{S}_{ia})$, where $\rho_{ia}$ is the density matrix for a single site's spin. These single-site density matrices are then treated as variational parameters, subject to the constraints of normalization (Tr$\{ \rho_{ia} \}$ = 1) and the order parameter definition (Tr$\{ \rho_{ia} \vect{S}_{ia} \} = \vect{m}_{ia}$). Enforcing these constraints using the sets of Lagrange multipliers $\{\xi_{ia}\}$ and $\{\vect{A}_{ia}\}$ yields
\begin{equation}
F(\{\rho_{ia}\},\{\xi_{ia}\},\{\vect{A}_{ia}\})= \textrm{Tr}\{\rho \mathcal{H}\} + T\textrm{Tr}\{\rho \ln \rho \} - T\textrm{Tr}\left\{\sum_{i,a} \xi_{ia} (\rho_{ia}-1)\right\} - T\textrm{Tr}\left\{\sum_{i,a} (\rho_{ia}\vect{S}_{ia} - \vect{m}_{ia}) \cdot \vect{A}_{ia}\right\}.\label{eq:FELagrange}
\end{equation}
Minimizing the free energy with respect to the variational parameters, the single-site density matrices are found to be $\rho_{ia} = \frac{1}{Z_{ia}} e^{\vect{A}_{ia} \cdot \vect{S}_{ia}}$. The partition function $Z_{ia}$ is computed by integrating over all spin configurations in spherical coordinates, given that the spins are treated classically (that is, as continuous vectors); this yields $Z_{ia} = \frac{4\pi}{A_{ia}} \sinh(A_{ia})$, where $A_{ia} = \abs{\vect{A}_{ia}}$. Using all of the above relations and computing the traces in Eq. \eqref{eq:FELagrange}, the free energy simplifies to
\begin{equation}
F = \frac{1}{2} \sum_{a,b} \sum_{i,j} \sum_{\mu, \nu} J_{ia,jb}^{\mu \nu}m_{ia}^{\mu}m_{jb}^{\nu} - \sum_{i,a} \vect{h}_{ia} \cdot \vect{m}_{ia} + \sum_{i,a} (\vect{H}_{ia} \cdot \vect{m}_{ia} - \frac{1}{\beta}\ln(Z_{ia})),
\end{equation}
where $\vect{H}_{ia} \equiv \frac{\vect{A}_{ia}}{\beta}$ can be considered as a local field. Explicitly, by minimizing the free energy with respect to the order parameters $\vect{m}_{ia}$, $\frac{\partial F}{\partial m_{ia}^{\mu}} = 0$, it can be shown that
\begin{equation}
H_{ia}^{\mu} = -\sum_{j,b,\nu} J_{ia,jb}^{\mu \nu} m_{jb}^{\nu} + h_{ia}^{\mu}. \label{eq:LocalField}
\end{equation}
Using this expression for $H_{ia}^{\mu}$, the free energy, $f$, averaged over all $N$ sites of the lattice is
\begin{equation}
f = \frac{F}{N} = \frac{X(m)}{N} - \frac{1}{N \beta} \sum_{i,a} \ln (Z_{ia}), \label{eq:VMFTFreeEnergy}
\end{equation}
where
\begin{equation}
X(m) \equiv -\frac{1}{2} \sum_{a,b} \sum_{i,j} \sum_{\mu, \nu} J_{ia,jb}^{\mu \nu}m_{ia}^{\mu}m_{jb}^{\nu}.
\end{equation}
Lastly, using the identity $\frac{\partial f}{\partial \vect{H}_{ia}} = -\vect{m}_{ia}$ for the order parameter at each site, one finds
\begin{equation}
\vect{m}_{ia} = \frac{\vect{H}_{ia}}{\abs{\vect{H}_{ia}}} \left[\coth{(\beta \abs{\vect{H}_{ia}})} -\frac{1}{\beta \abs{\vect{H}_{ia}}}\right]. \label{eq:MFTSelfCon}
\end{equation}
This equation relates the local field $\vect{H}_{ia}$ with the order parameter $\vect{m}_{ia}$ at each site, which can be calculated self-consistently.

The evolution of a chosen Palmer-Chalker state (e.g. $\langle xy \rangle$) in the presence of an applied magnetic field  ${\bm B}$ is accomplished within variational mean field theory as follows. For a given magnitude and direction of the field (and at zero temperature), the tetrahedra of the pyrochlore lattice are initiated with an initial spin configuration. At zero-field, this initial spin configuration is that of the chosen Palmer-Chalker state. At finite field, this initial spin configuration is the mean field solution from the previous field magnitude. For example, if the mean field calculation is done for $B$ values incremented by 0.01 T, the initial configuration used for $B = 0.11$ T is the mean field solution from $B = 0.10$ T. In this way, the evolution of the chosen zero-field Palmer-Chalker state can be tracked as a function of $B$.  The self-consistency equation Eq.~\eqref{eq:MFTSelfCon} is then solved iteratively for zero temperature until convergence is attained. The resulting spin configuration is denoted with a subscript $h$ (e.g. $\langle xy \rangle_h$) to represent that it is the field-evolved configuration of the originally-chosen Palmer-Chalker state. This field-evolved Palmer-Chalker (FEPC) state for a given $B$ can then be used to tile the pyrochlore lattice and solve the self-consistency equation for \textit{finite} temperatures for the same $B$ value, showing how the chosen Palmer-Chalker state evolves with temperature as well. This process is followed for \textit{each} of the six parent $B=0$ Palmer-Chalker states, in order to track their field- and temperature-evolution individually.

\begin{figure}[ht!]
    \subfigure[]{\includegraphics[scale=\BCscalefactor]{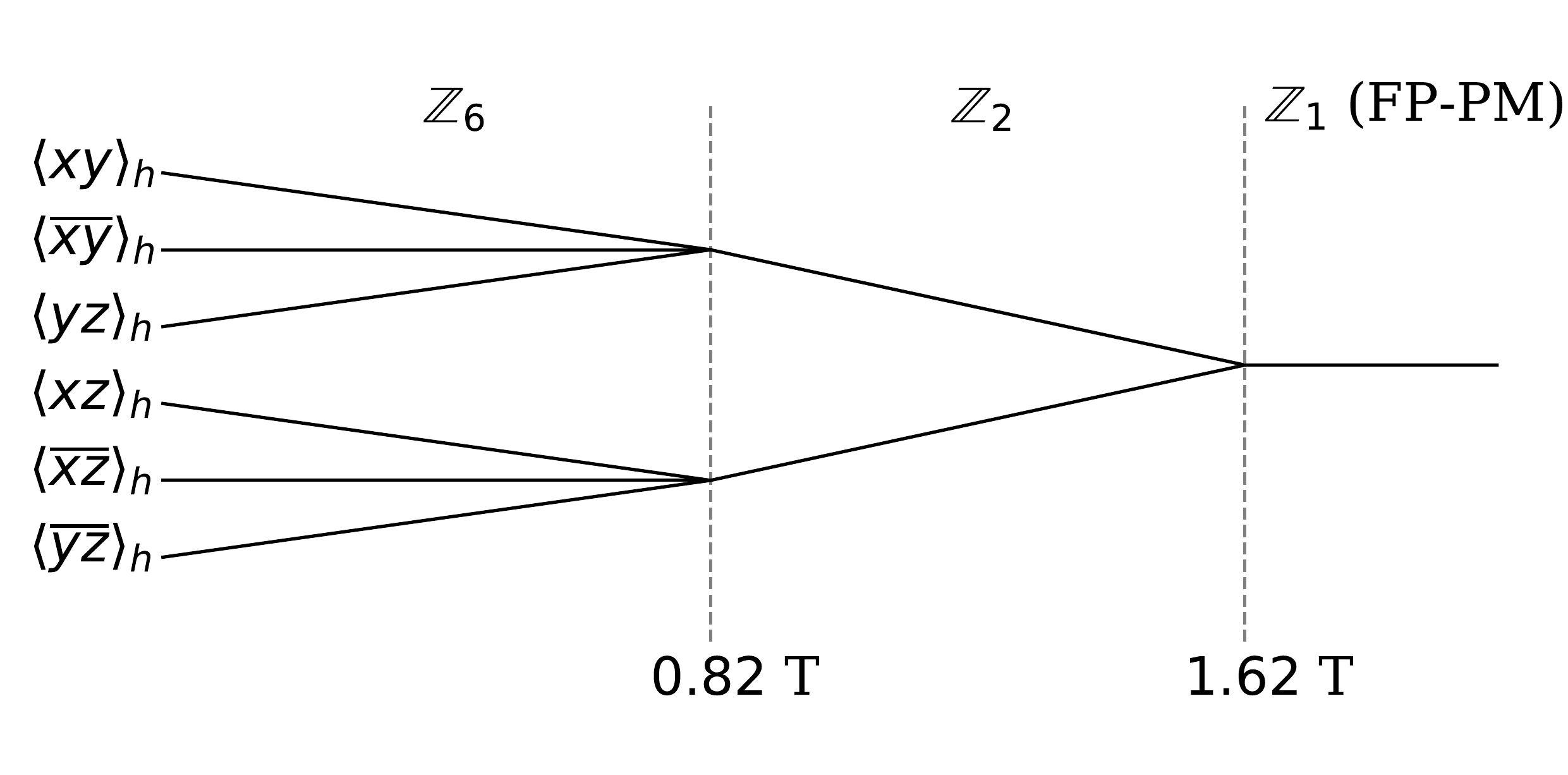}} \label{fig:Branches100}
    \hspace*{\fill}
    \subfigure[]{\includegraphics[scale=\MFTscalefactor]{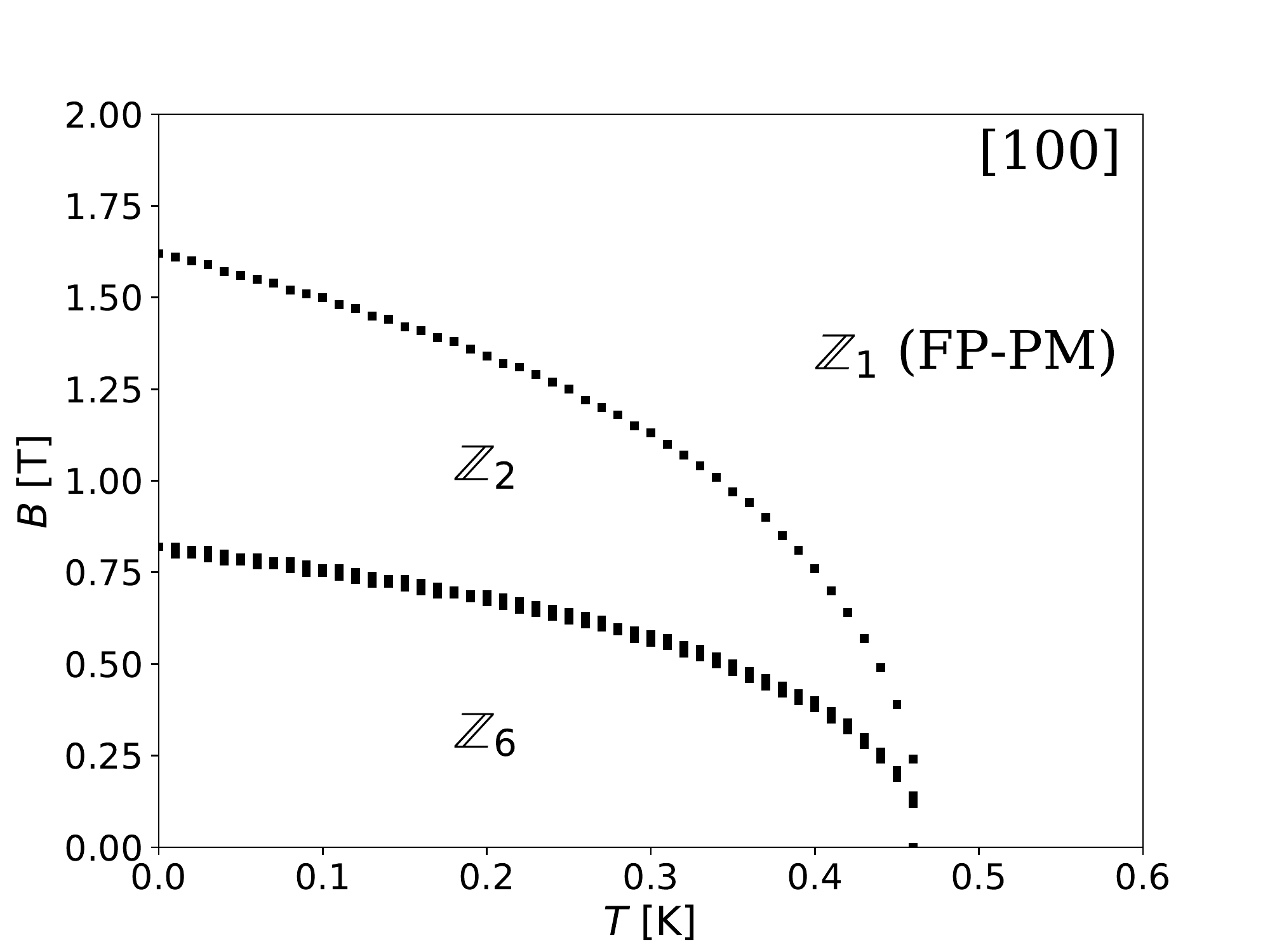}} \label{fig:VMFT100}
    
    \subfigure[]{\includegraphics[scale=\BCscalefactor]{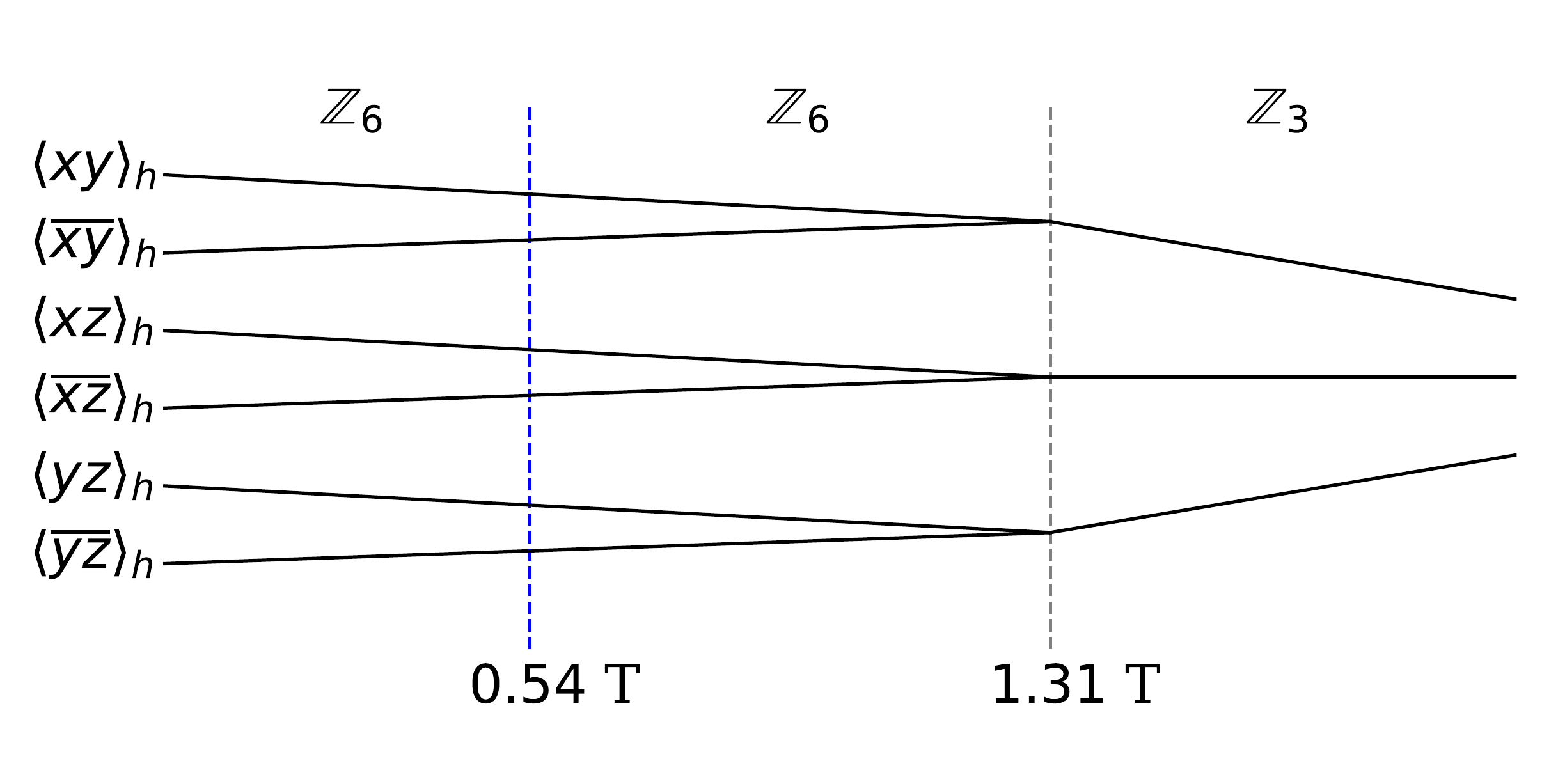}} \label{fig:Branches111}
    \hspace*{\fill}
    \subfigure[]{\includegraphics[scale=\MFTscalefactor]{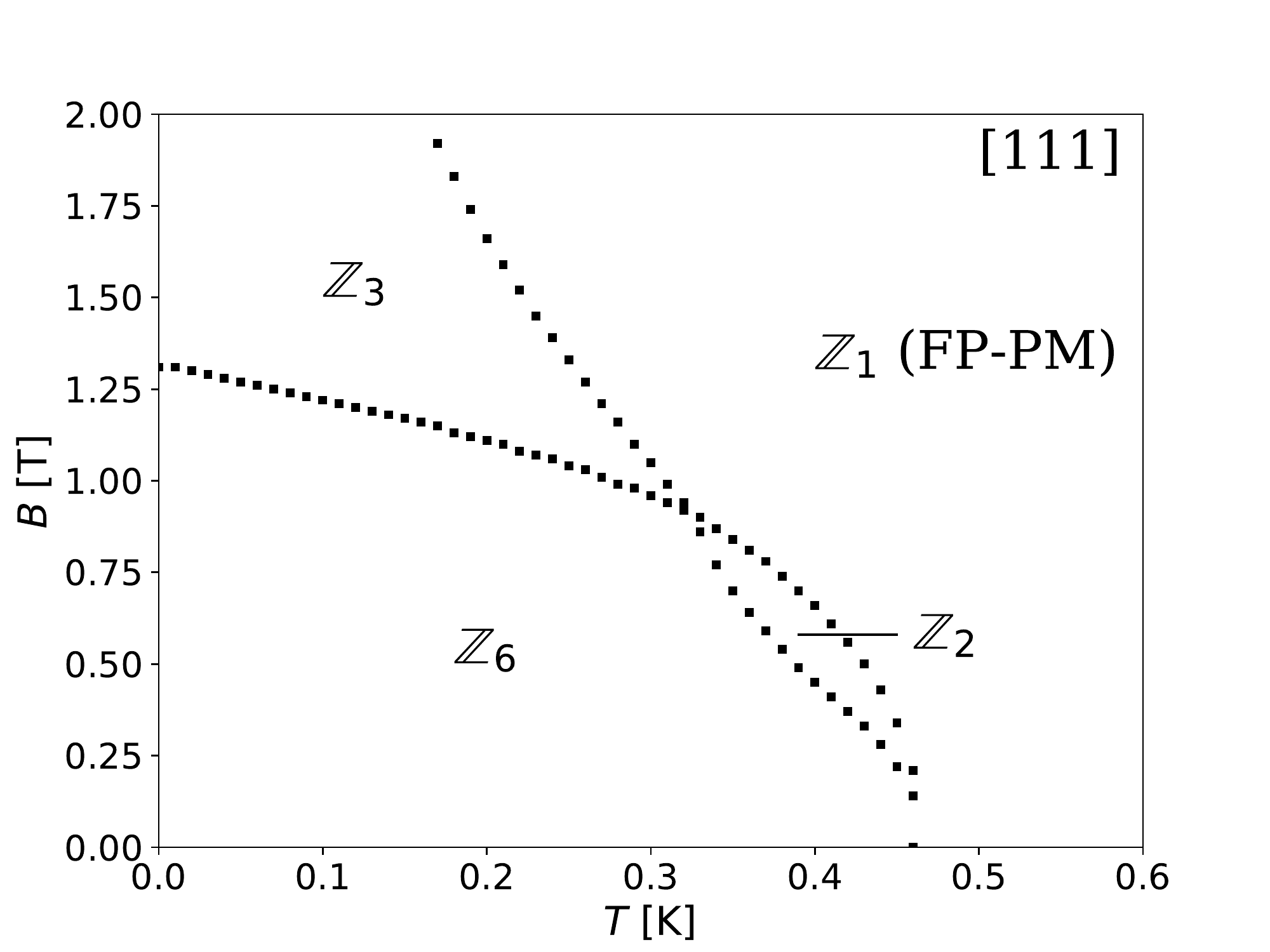}}\label{fig:VMFT111}
    
    \subfigure[]{\includegraphics[scale=\BCscalefactor]{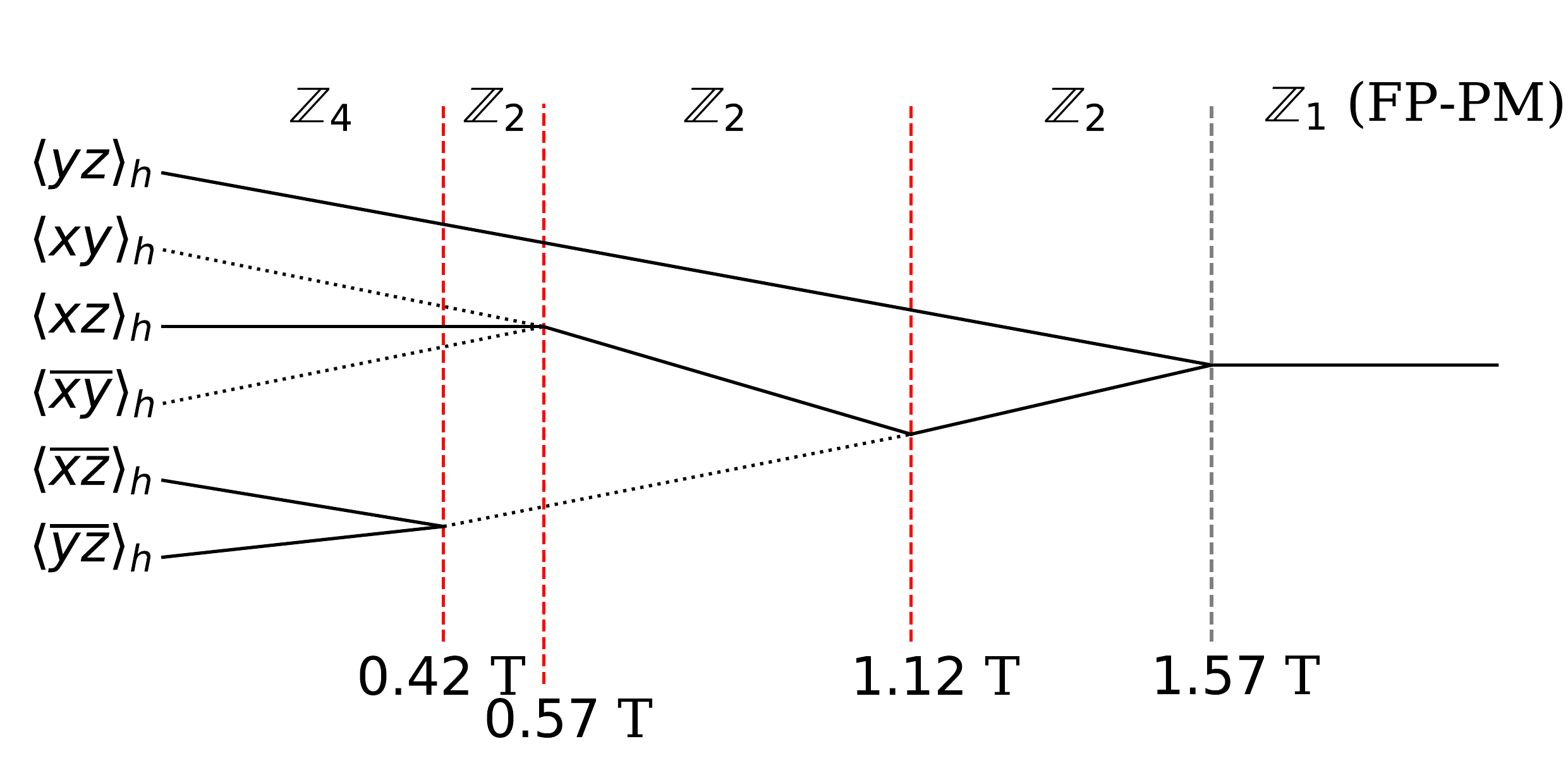}} \label{fig:Branches110}
    \hspace*{\fill}
    \subfigure[]{\includegraphics[scale=\MFTscalefactor]{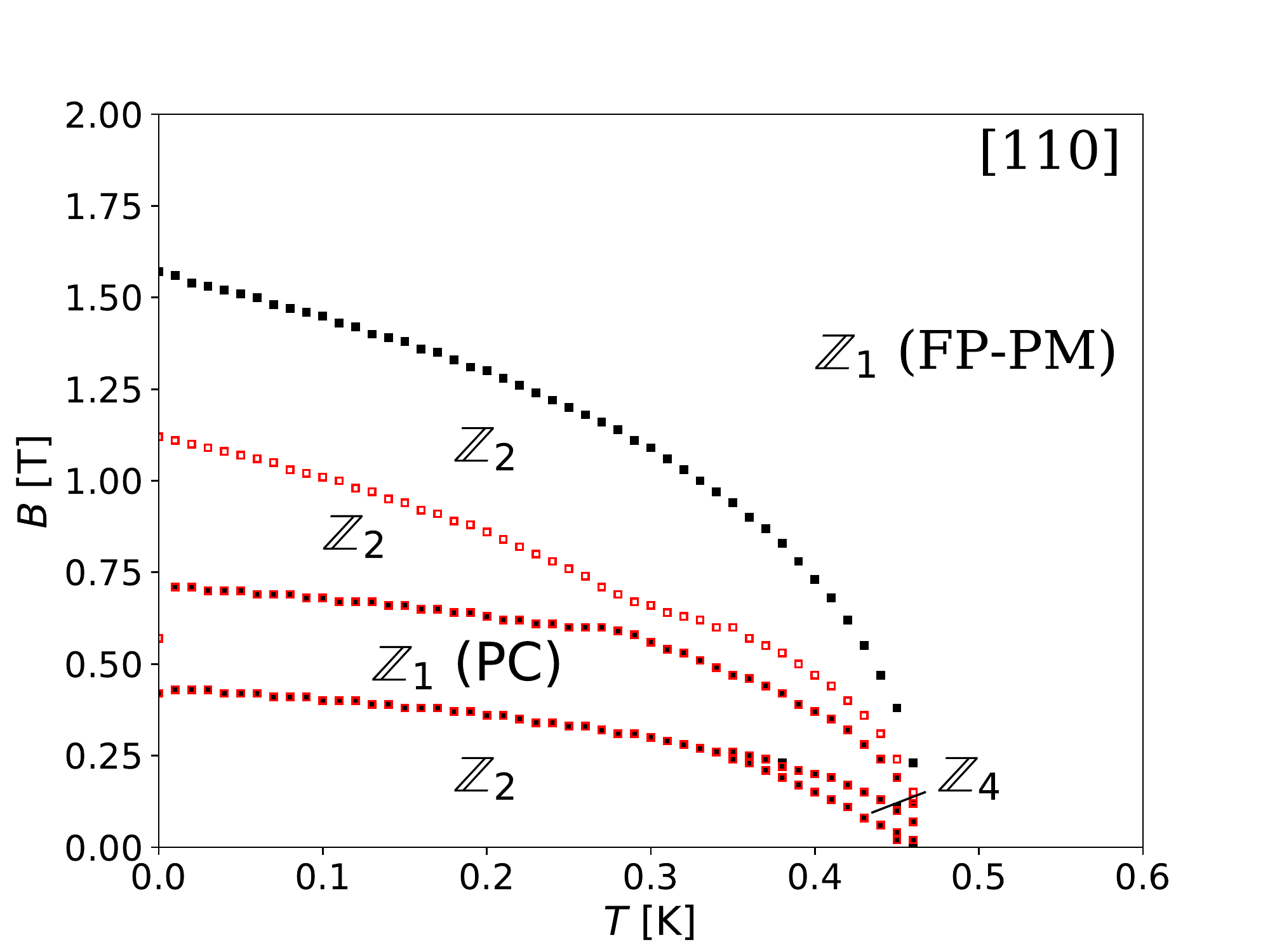}} \label{fig:VMFT110}
    
    \caption{Zero temperature representations of field-evolved Palmer-Chalker (FEPC) mergers ((a), (c), and (e)) and mean field $B$-$T$ phase diagrams ((b), (d), and (f)) for the [100], [111], and [110] field directions, respectively. In the FEPC merger diagrams, vertical grey lines represent critical fields of the merging transitions; the vertical blue line represents field-induced selection of the $\psi_2$ phase out of the accidentally-degenerate $\Gamma_5$ manifold; vertical red lines represent changes in the set of degenerate FEPC states. Solid lines denote the FEPC states that are degenerate and minimize the energy; dotted lines represent FEPC states that are excited and do not minimize the energy. In the mean field phase diagrams, filled black squares represent a phase transition to a phase of different discrete symmetry due to merging; open red squares represent a change in the set of degenerate FEPC states. In both sets of diagrams, $\mathbb{Z}_n$ denotes the discrete symmetry of the phase, which has $n$ distinct spin configurations out of the degenerate FEPC states. FP-PM denotes the field-polarized paramagnetic phase; PC denotes Palmer-Chalker order. Note that the zero temperature phases in the [110] field direction differ from the finite temperature phases due to entropic effects, for reasons discussed further in Section \ref{SM-Analysis110MC}.}
    \label{fig:VMFTResults}
\end{figure}

The mean field calculations are performed for a pyrochlore lattice of size $L=2$ (in terms of cubic unit cells) for a total of $16L^3 = 128$ sites. At all fields and temperatures, the spin configuration is always found to be $\vect{q} = 0$ ordering wave vectors. The phases can be labelled by (i) how many of the six FEPC states minimize the free energy (which may be less than six), and (ii) how many of these degenerate FEPC states have distinct spin configurations. If there are $n$ distinct spin configurations out of all degenerate FEPC states, that phase is labelled as a $\mathbb{Z}_n$ phase, reflecting the order of the discrete degeneracy. Phase transitions are then indicated by a reduction in this discrete symmetry as the field increases, referred to as ``merger'' transitions. The mean field phase diagram for the $[100]$, $[111]$, and $[110]$ field directions, as well as a pictorial representation of the merger transitions at zero temperature, are illustrated in Fig.~\ref{fig:VMFTResults}. 

Variational mean field theory including dipolar interactions has also been performed by use of the Ewald summation method~[\onlinecite{EwaldSum, Oitmaa13, VMFT03}]. When including the dipolar interactions, the $g$-tensor must be properly incorporated to map the interaction (between dipoles) onto the pseudospin $S=\frac{1}{2}$ representation. As well, it should be noted that the nearest-neighbor part of the dipolar interaction is already included in the nearest-neighbor exchange couplings $(J_1, J_2, J_3, J_4)$. With the inclusion of dipolar interactions, the critical temperatures and fields are reduced relative to Fig.~\ref{fig:VMFTResults}, but the $\vect{q} = 0$ orders, $\mathbb{Z}_n$ phases, and overall topology of the phase diagrams remain unchanged.

\clearpage
\section{Classical Spin-Wave Expansion}
\label{SM-TechSW}

A classical spin-wave expansion for $\vect{q} = 0$ spin ordering on the pyrochlore lattice and without an applied field is detailed in Ref. [\onlinecite{MPCPyrochlore}]; the derivation below for a classical spin-wave expansion in a finite field is similar. Informed by the mean field results at zero temperature, $\vect{q} = 0$ ordering is assumed and dipolar interactions, which do not change the topology of the phase diagram, are excluded. Under these assumptions, the Hamiltonian is
\begin{equation}
\mathcal{H} = \frac{1}{2}\sum_a\sum_{ij}J_{ij}^{\mu \nu} S_{ia}^\mu S_{ja}^\nu - \mu_{\textrm B} \sum_a\sum_i g_i^{\mu \nu} S_{ia}^\mu B^\nu,    
\end{equation}
where $J_{ij}^{\mu \nu}$ is now restricted to only nearest neighbors, and the tetrahedra indices are suppressed because of the $\vect{q} = 0$ assumption. 

Starting with the classically-ordered ground state spin configuration on a single tetrahedron at $T = 0$, define the local coordinate system at each sublattice with $\{ \vect{u}_i, \vect{v}_i, \vect{w}_i \}$, where the classical spin vector is of length $S$ and points along $\vect{w}_i$. The other two local unit vectors are defined arbitrarily, so long as mutual orthogonality is satisfied. The fluctuations about this ordered spin vector, on sublattice  $i$ and tetrahedron $a$, can then be expressed as
\begin{align}
    \vect{S}_{ia} = 
    \begin{bmatrix}
    \sqrt{S} \delta u_{ia} \\
    \sqrt{S} \delta v_{ia} \\
    \sqrt{S^2 - S\delta u_{ia}^2 - S\delta v_{ia}^2}
    \end{bmatrix} 
    \approx  
    \begin{bmatrix}
    \sqrt{S} \delta u_{ia} \\
    \sqrt{S} \delta v_{ia} \\
    S - \frac{1}{2}\delta u_{ia}^2 - \frac{1}{2}\delta v_{ia}^2
    \end{bmatrix}.
\end{align}
\noindent Substituting this perturbed spin into the original Hamiltonian and expanding to quadratic order in the fluctuations yields
\begin{equation}
    \begin{split}
        \mathcal{H} &= \frac{1}{2} N_\textrm{t} \sum_{ij} S^2 (\vect{w}_i \cdot \matr{J}_{ij} \cdot \vect{w}_j) 
        - \mu_{\textrm B} S N_\textrm{t} \sum_i (\vect{w}_i \cdot \matr{g}_{i} \cdot \vect{B}) 
        + \frac{S}{2} \sum_a \sum_{ij} \delta u_{ia} \delta u_{ja} (\vect{u}_i \cdot \matr{J}_{ij} \cdot \vect{u}_j) \\
        & + \delta u_{ia} \delta v_{ja} (\vect{u}_i \cdot \matr{J}_{ij} \cdot \vect{v}_j) 
        + \delta v_{ia} \delta u_{ja} (\vect{v}_i \cdot \matr{J}_{ij} \cdot \vect{u}_j)
        + \delta v_{ia} \delta v_{ja} (\vect{v}_i \cdot \matr{J}_{ij} \cdot \vect{v}_j) \\
        & -\frac{1}{2}(\delta u_{ia}^2 + \delta v_{ia}^2 + \delta u_{ja}^2 + \delta v_{ja}^2)(\vect{w}_i \cdot \matr{J}_{ij} \cdot \vect{w}_j) + \frac{1}{2}\mu_B \sum_a \sum_{ij} (\delta u_{ia}^2 + \delta v_{ia}^2) (\vect{w}_i \cdot \matr{g}_{i} \cdot \vect{B}).
    \end{split}
\end{equation}
\noindent The first two terms just represent the ground state energy, $\varepsilon_0$, where $N_\textrm{t}$ is the number of tetrahedra in the system ($N_\textrm{t}=\frac{N}{4}$, where $N$ is the number of spins in the system). The rest of the expression encapsulates the effect of fluctuations from exchange and Zeeman interactions.

Performing a Fourier transform over the reciprocal lattice vectors $\vect{q}$ of the FCC lattice and defining $\vect{u}(\vect{q}) \equiv (\delta u_1(\vect{q}), \delta u_2(\vect{q}), \delta u_3(\vect{q}), \delta u_4(\vect{q}), \delta v_1(\vect{q}), \delta v_2(\vect{q}), \delta v_3(\vect{q}), \delta v_4(\vect{q}))$, the fluctuation contribution to the Hamiltonian can be written as:
\begin{equation}
    \mathcal{H} = \varepsilon_0 + \frac{1}{2} \sum_{\vect{q}} \vect{u}(-\vect{q}) \left(\matr{M}(\vect{q}) + \matr{N}(\vect{q})\right) \vect{u}(\vect{q}).
\end{equation}
\noindent The matrices $\matr{M}(\vect{q})$ and $\matr{N}(\vect{q})$ can be written in block matrix form, composed of four separate $4 \times 4$ blocks. They are:
\begin{equation}
    \matr{M} = 2S\begin{bmatrix}
    \matr{M}^{11}(\vect{q}) & \matr{M}^{12}(\vect{q}) \\
    \matr{M}^{21}(\vect{q}) & \matr{M}^{22}(\vect{q}) 
    \end{bmatrix}
\end{equation}
\begin{align}
    M_{ij}^{11} &= \cos{(\vect{q} \cdot \vect{r}_{ij})}(\vect{u}_i \cdot \matr{J}_{ij} \cdot \vect{u}_j - \delta_{ij} \sum_l \vect{w}_l \cdot \matr{J}_{lj} \cdot \vect{w}_j) \nonumber \\
    M_{ij}^{12} &= \cos{(\vect{q} \cdot \vect{r}_{ij})}(\vect{u}_i \cdot \matr{J}_{ij} \cdot \vect{v}_j) \nonumber \\
    M_{ij}^{21} &= \cos{(\vect{q} \cdot \vect{r}_{ij})}(\vect{v}_i \cdot \matr{J}_{ij} \cdot \vect{u}_j) \nonumber \\
    M_{ij}^{22} &= \cos{(\vect{q} \cdot \vect{r}_{ij})}(\vect{v}_i \cdot \matr{J}_{ij} \cdot \vect{v}_j - \delta_{ij} \sum_l \vect{w}_l \cdot \matr{J}_{lj} \cdot \vect{w}_j) \nonumber
\end{align}
\begin{equation}
    \matr{N} = \mu_{\textrm B}\begin{bmatrix}
    \matr{N}^{11}(\vect{q}) & 0 \\
    0 & \matr{N}^{22}(\vect{q}) 
    \end{bmatrix}
\end{equation}
\begin{align}
    N_{ij}^{11} = N_{ij}^{22} = \delta_{ij} \vect{w}_i \cdot \matr{g}_{i} \cdot \vect{B} \nonumber.
\end{align}
Diagonalizing this harmonic spin-wave Hamiltonian yields the classical spin-wave dispersions $\kappa_{\nu}(\vect{q})$ as a function of the wavevector $\vect{q}$. Since the classical spin-wave Hamiltonian is quadratic, the partition function and free energy of the classical spin-waves can be calculated exactly. Assuming the above diagonalization has been performed to find the normal modes $\phi_{\nu}(\vect{q})$ and dispersions $\kappa_{\nu}(\vect{q})$, the partition function and free energy are given by
\begin{align}
    Z &= \int \left[ \prod_{\vect{q}} \prod_{\nu=1}^8 d\phi_\nu(\vect{q}) \right] e^{-\frac{1}{T}\left(\varepsilon_0 + \frac{1}{2}\sum_{\vect{q}}\sum_{\nu=1}^8 \kappa_{\nu}(\vect{q}) \phi_\nu(\vect{q})\phi_\nu(-\vect{q})\right)} \\
    &= e^{-\frac{\varepsilon_0}{T}} \prod_{\vect{q}} \prod_{\nu=1}^8 \sqrt{\frac{2\pi T}{\kappa_{\nu}(\vect{q})}} \\
    \implies F &= \varepsilon_0 + \frac{T}{2}\sum_{\vect{q}} \sum_{\nu=1}^8 \ln(\kappa_{\nu}(\vect{q})) - NT \ln(2\pi T).
\end{align}
\noindent If there are $N_q = \frac{N}{4}$ wavevectors in the sum, then the free energy per spin is
\begin{equation}
    f = \frac{\varepsilon_0}{N} + \frac{T}{8N_q}\sum_{\vect{q}} \sum_{\nu=1}^8 \ln(\kappa_{\nu}(\vect{q})) - T \ln(2\pi T).
\end{equation}
Note that the entropy per spin $s$ can be calculated using $s = -\frac{\partial f}{\partial T}$. Computing this, but keeping only the terms that depend on the spin-wave dispersions $\kappa_\nu(\vect{q})$, yields
\begin{equation}
    s = -\frac{1}{8N_q}\sum_{\vect{q}} \sum_{\nu=1}^8 \ln(\kappa_{\nu}(\vect{q})).
\end{equation}
The dispersions $\kappa_{\nu}(\vect{q})$ for each of the FEPC states are shown below at various relevant choices of the applied field (e.g. at the merger transitions). 
\begin{figure}[ht!]
    \subfigure[]{\includegraphics[scale=\CSWscalefactor]{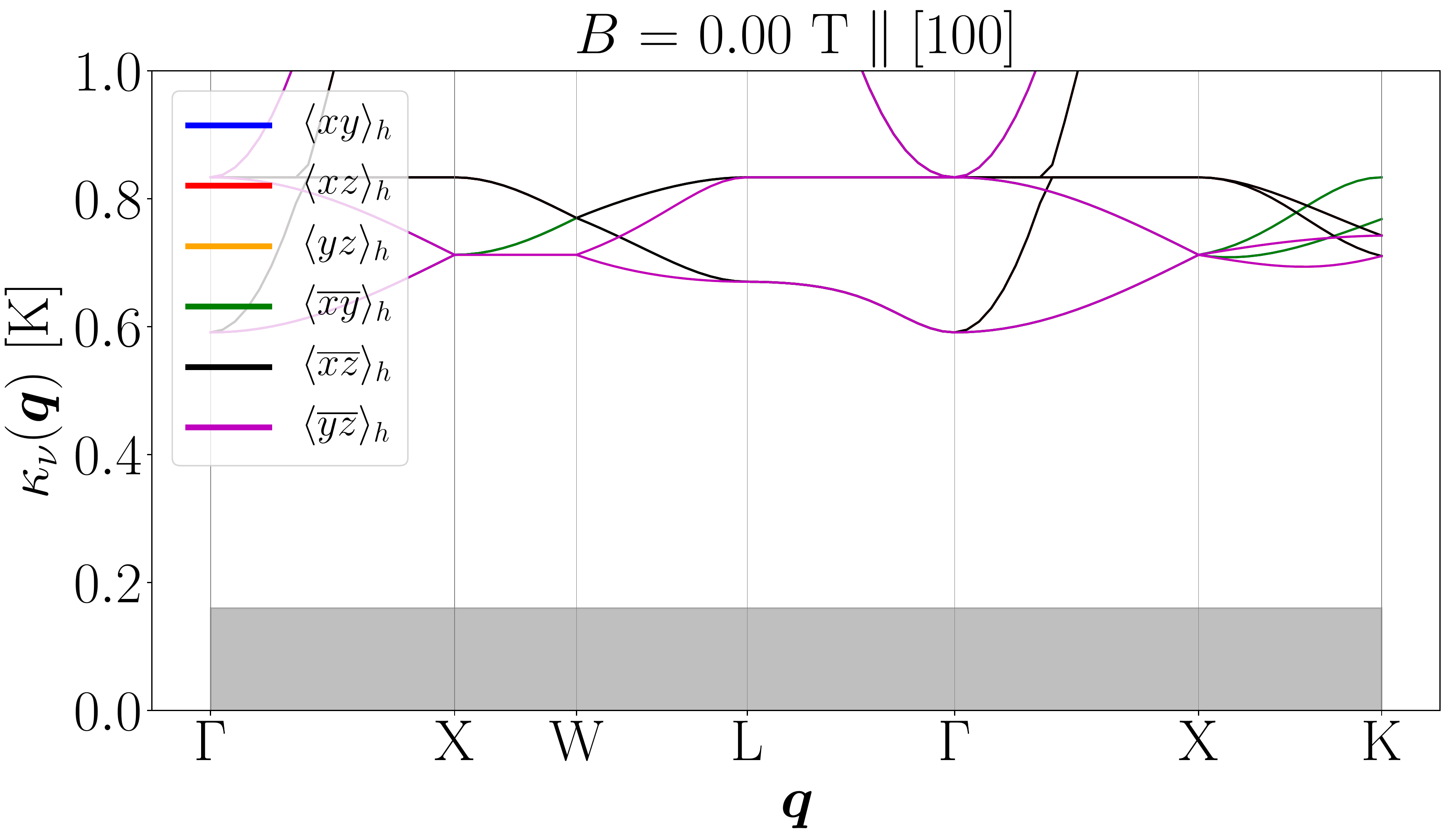}} \label{fig:CSW100_000}
    \subfigure[]{\includegraphics[scale=\CSWscalefactor]{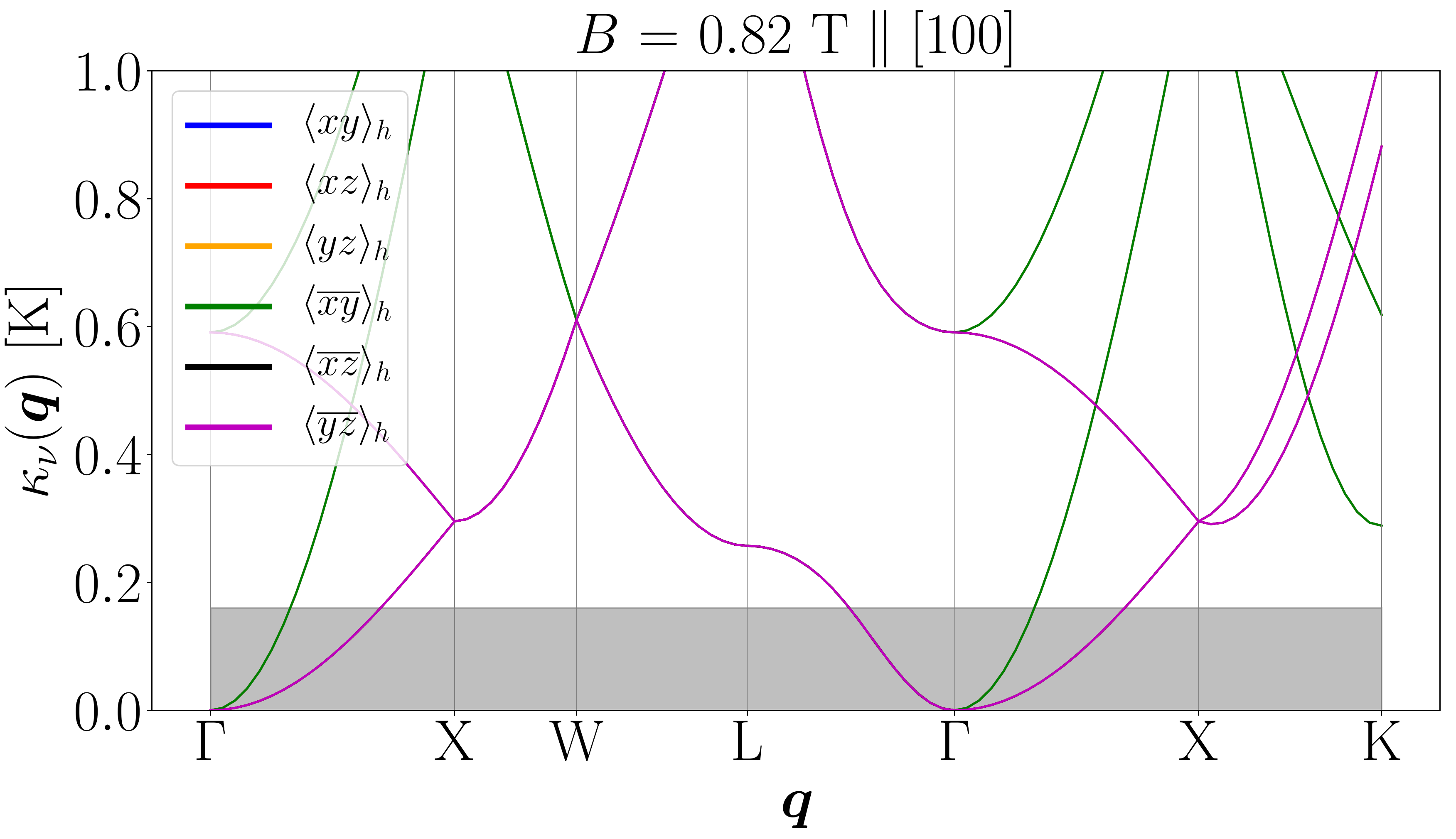}} \label{fig:CSW100_082}
    \subfigure[]{\includegraphics[scale=\CSWscalefactor]{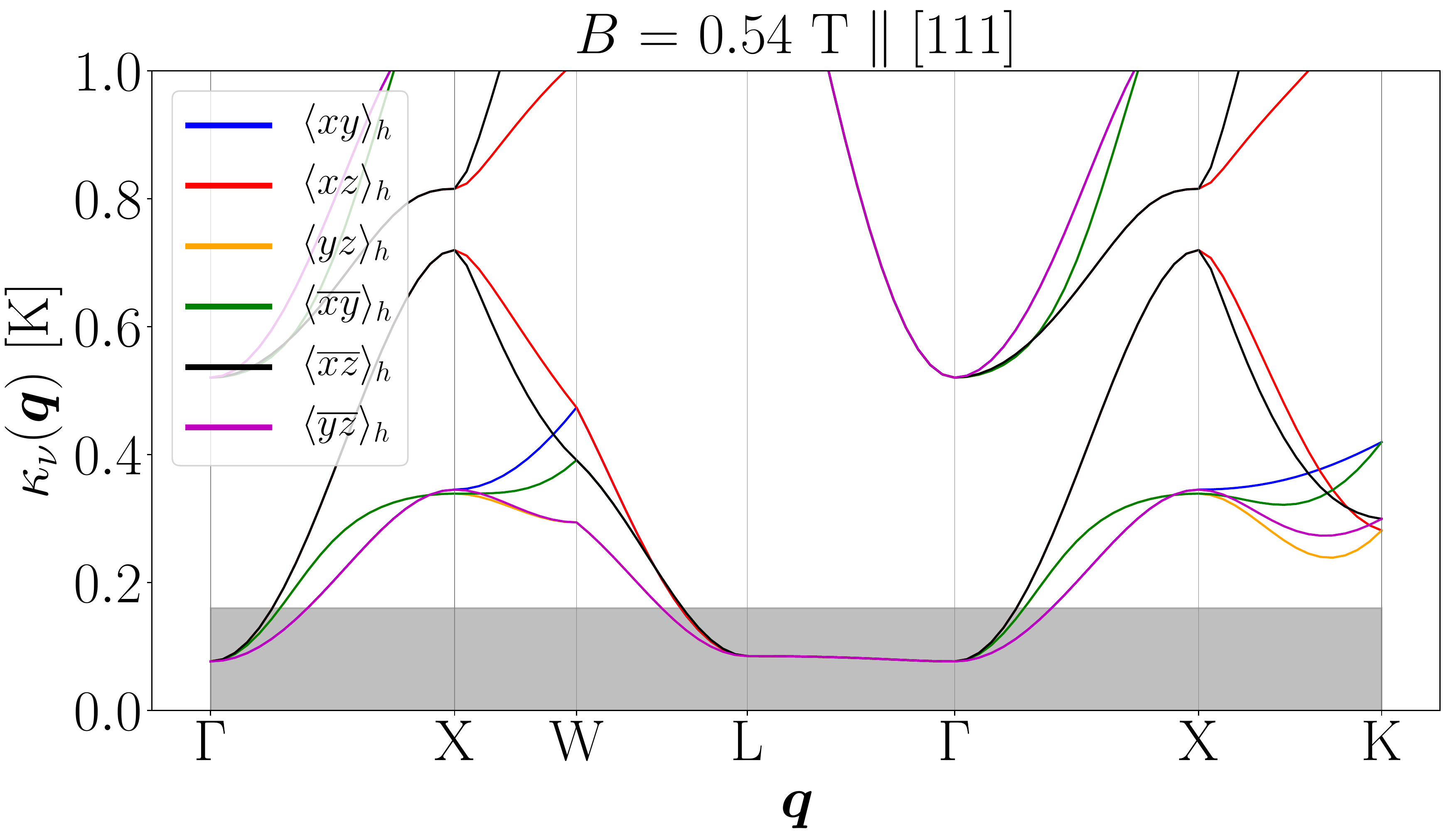}} \label{fig:CSW111_054}
    \subfigure[]{\includegraphics[scale=\CSWscalefactor]{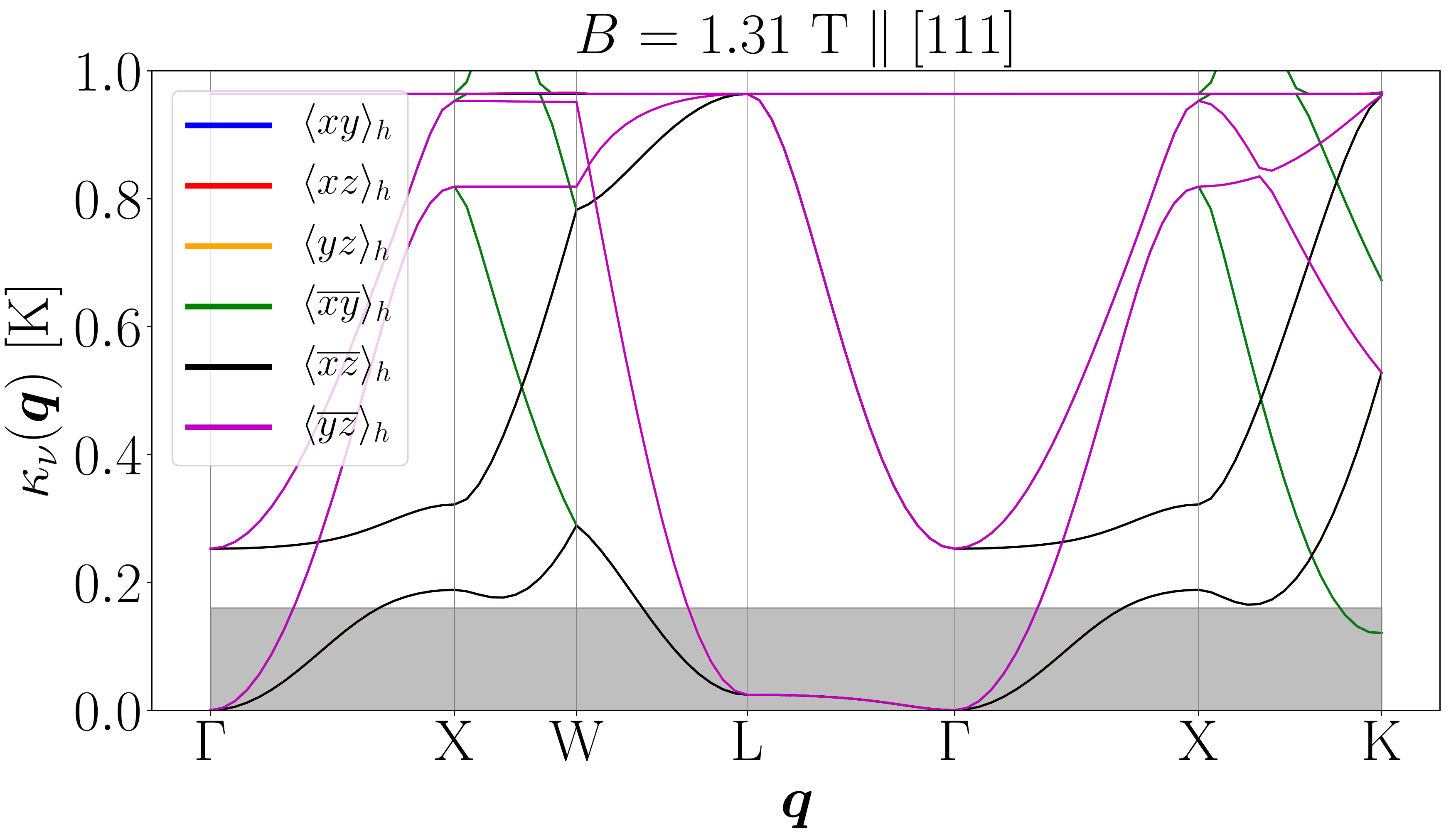}} \label{fig:CSW111_131}
    \subfigure[]{\includegraphics[scale=\CSWscalefactor]{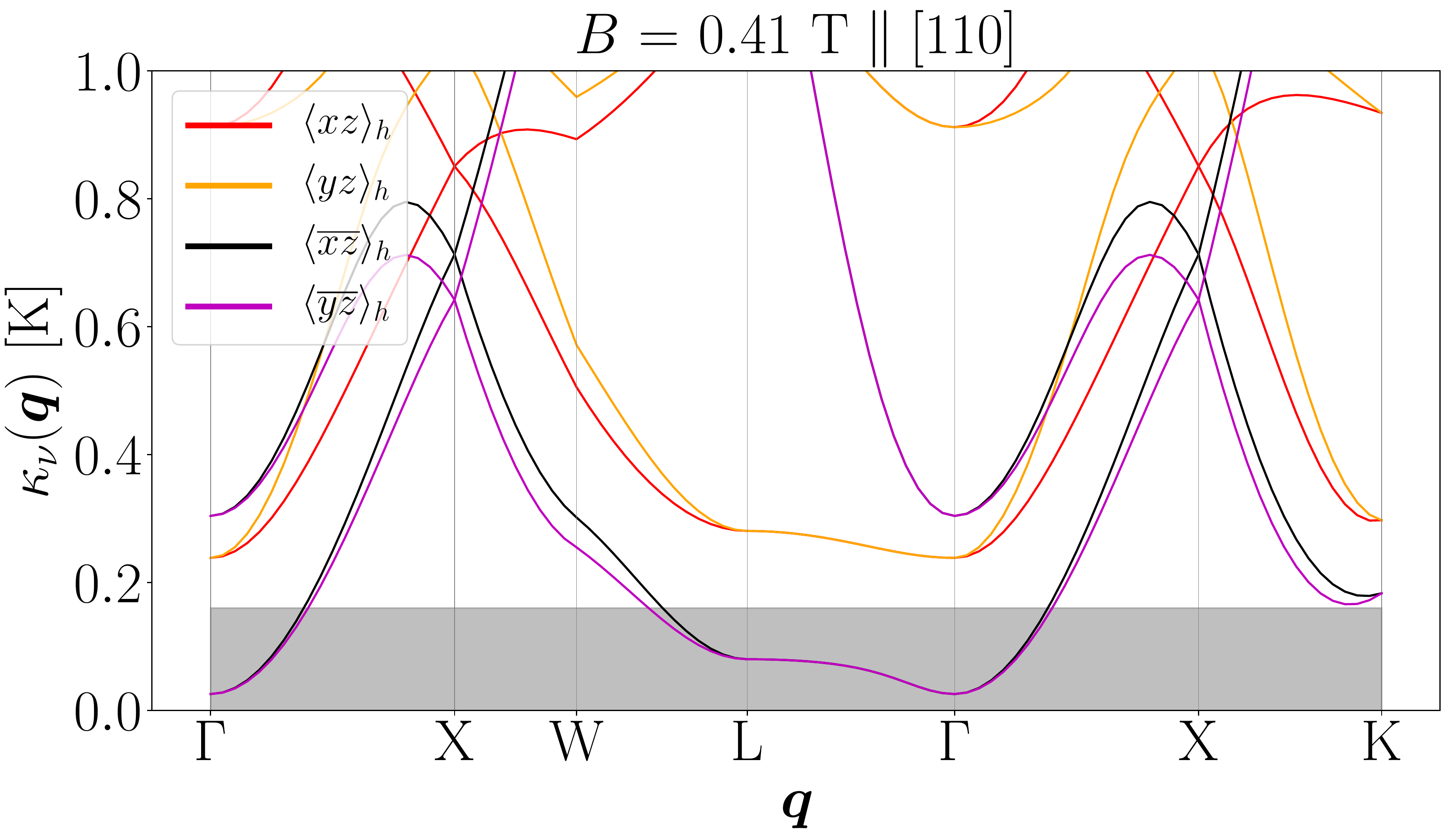}} \label{fig:CSW110_041}
    \subfigure[]{\includegraphics[scale=\CSWscalefactor]{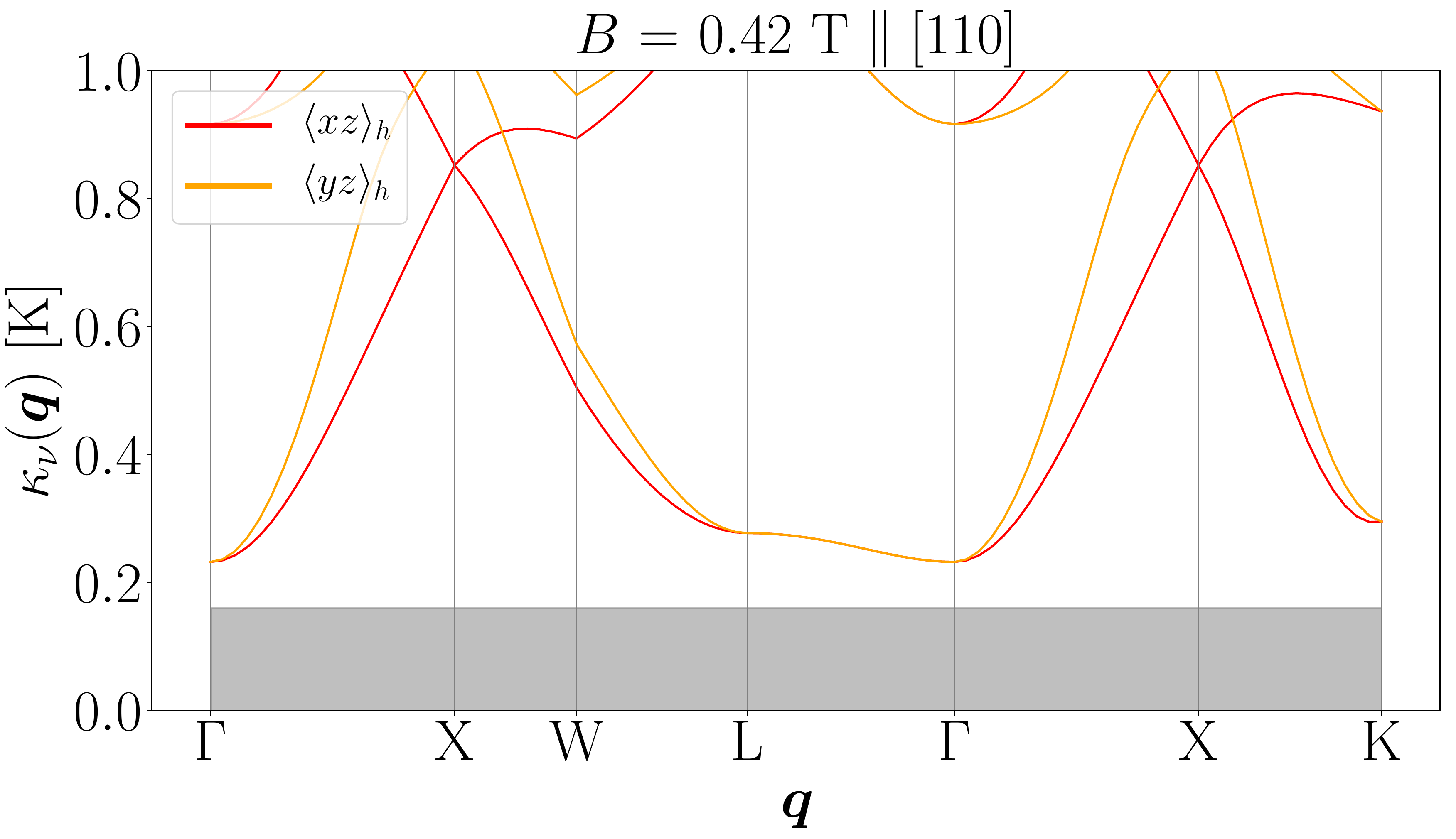}} \label{fig:CSW110_042}
    \subfigure[]{\includegraphics[scale=\CSWscalefactor]{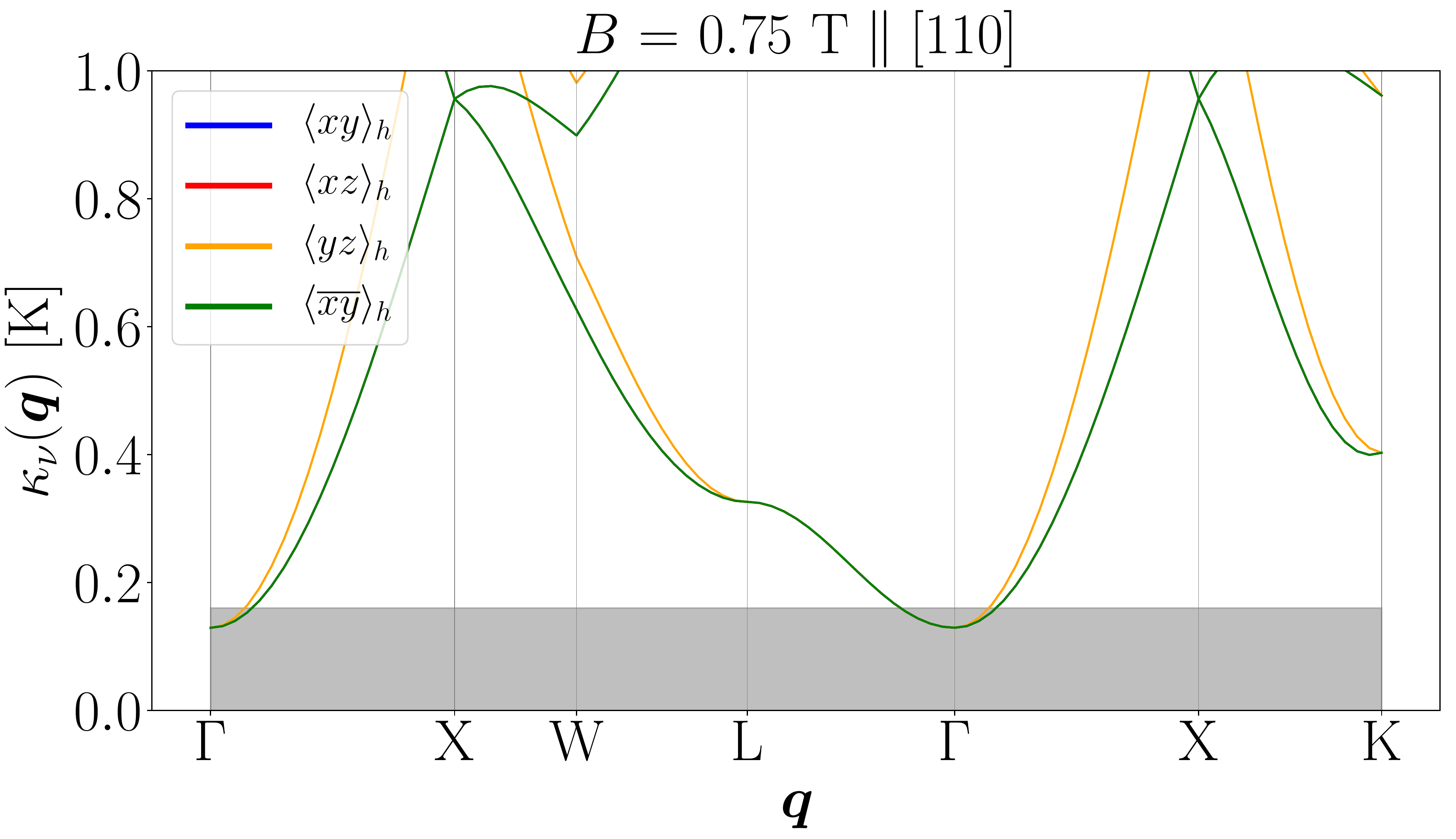}} \label{fig:CSW110_075}
    
    \caption{Classical spin-wave dispersions for all Palmer-Chalker states that are degenerate in their energy, for a given field direction and magnitude. The shaded grey box represents energy scales below the zero-field Monte Carlo critical temperature of $T_c \approx 160$ mK. The chosen wavevectors are taken from the Brillouin zone of the FCC lattice. Note that some curves may overlap at high-symmetry points or due to their degeneracies in a field (e.g. $\langle xy \rangle_h$ and $\langle \overline{xy} \rangle_h$ for $B \geq 0.82$ T along the [100] direction).} \label{fig:CSWPlots}
\end{figure}

\clearpage
\section{Monte Carlo simulations and Parameters}
\label{SM-MC}

Monte Carlo simulations are performed on systems of classical Heisenberg spins with $N=16L^{3}$ sites, where $L^{3}$ is the number of cubic unit cells. The spin length is $|S|=1/2$. Several update algorithms are used together: the heatbath method, over-relaxation and parallel tempering. Parallel tempering is done every 100 Monte Carlo steps (MCS) and overrelaxation is done at every MCS. Thermalization is made in two steps: first a slow annealing from high temperature to the temperature of measurement $T$ during $t_{e}$ MCS followed by $t_{e}$ MCS at temperature $T$. After thermalization, measurements are done every 10 MCS during $t_{m}=10~t_{e}$ MCS.

The characteristics of our simulations are typically:
\begin{itemize}
    \item $4 < L < 10$,
    \item $10^6 \leq t_{m} \leq 5. 10^7$ MCS,
    \item 100 different temperatures (regularly spaced) between 0 and 200 mK for parallel tempering.
\end{itemize}

Please note that different exchange parameters for \ESO\ have been investigated, within the error bars of  Ref.~[\onlinecite{ESOPetitPRL}]. In particular, a systematic search in the [111] direction has shown a consistent increase of the transition temperature between 0 and 0.4 T, confirming the robustness of the reentrance in this exchange-parameter region.

\clearpage
\section{Analysis of the [110] Results}
\label{SM-Analysis110MC}

Here we discuss the Monte Carlo phase diagram in a [110] field (Fig. 2(b) of the main text). There are two occurrences of reentrance, as well as a pronounced dip in $T_{\rm c}$ that occurs between them. As with the [100] and [111] phase diagrams, these three features can also be understood as originating from soft modes that arise at the merging of $T=0$ FEPC states. In this field and at $T=0$, the six FEPC states do not all minimize the energy (for reasons discussed later on in this section). Therefore, when determining the merging of FEPC states, we primarily consider those ground states which simultaneously minimize the energy (as shown in Fig. \ref{fig:VMFTResults}(e)), since these are the states the system can be found in. Hence, as $B$ is increased from $0$ and approaches a merging transition (between $\langle \overline{xz} \rangle_h$ and $\langle \overline{yz} \rangle_h$) around $B \approx 0.4$ T, soft modes arise (see Fig. \ref{fig:CSWPlots}(e)) and increase the thermal fluctuations within the ordered phase, producing the first (lower) reentrance. As $B$ continues to increase to and above $B \approx 0.42$ T, however, the merging FEPC states that provided these soft modes (namely, $\langle \overline{xz} \rangle_h$ and $\langle \overline{yz} \rangle_h$) are no longer energetically preferred (see Fig. \ref{fig:VMFTResults}(e) and Fig. \ref{fig:CSWPlots}(f)). These soft modes are therefore removed from the system, taking away their entropic support and producing a dip in $T_{\rm c}$. As $B$ is increased further and enters the region $0.57 {\rm \ T} < B < 1.12 {\rm \ T}$, soft modes still arise (see Fig. \ref{fig:CSWPlots}(g)), although there are no merger transitions between FEPC ground states, only between ground and excited states (as shown in Fig. \ref{fig:VMFTResults}(e)). These soft modes give rise to the second (upper) reentrance. Since they are sustained for a range of $B$, the upper reentrant lobe is broad in the $B$ direction.


Although the mechanisms of reentrance are similar to the other two directions, the mean field theory results for the [110] field direction do stand out among the three field directions we studied, given that the zero temperature phases differ from those at finite temperature. The objective of the remainder of this section is to clarify this difference.

Consider very low temperatures $T=\epsilon$ where $\epsilon \to 0^+$. When $T=0$ exactly, Eq. \eqref{eq:VMFTFreeEnergy} reduces to just the energy, whereas Eq. \eqref{eq:MFTSelfCon} implies that $\vect{m}_{ia} = \frac{\vect{H}_{ia}}{|\vect{H}_{ia}|}$. For $T=\epsilon$, $\vect{m}_{ia} \approx \frac{\vect{H}_{ia}}{|\vect{H}_{ia}|}$. As well, the free energy now gains a contribution from the entropy, namely $- \frac{1}{N\beta}\sum_{i,a}\ln (Z_{ia})$. Using $A_{ia} = \beta |\vect{H}_{ia}|$ in the partition function:
\begin{align}
    - \frac{1}{N\beta}\sum_{i,a}\ln (Z_{ia}) &= - \frac{1}{N\beta} \sum_{i,a} \ln(\frac{4\pi}{\beta |\vect{H}_{ia}|} \sinh(\beta |\vect{H}_{ia}|)) \nonumber \\
    &= - \frac{\epsilon}{N} \sum_{i,a} \ln(\frac{4\pi \epsilon}{ |\vect{H}_{ia}|}) + \ln (\sinh(\frac{|\vect{H}_{ia}|}{\epsilon})). \nonumber
\end{align}
As $\epsilon \to 0^+$, $\epsilon \ln(\epsilon) \to 0$ as well. Hence, the first term in the above summation is not an important contribution to the entropy at $T=0^+$. On the other hand, $\sinh(\frac{|\vect{H}_{ia}|}{\epsilon})$ scales roughly as $e^{\frac{|\vect{H}_{ia}|}{\epsilon}}$ for $\epsilon \to 0^+$. Hence:
\begin{align}
    - \frac{1}{N\beta}\sum_{i,a}\ln (Z_{ia}) &\approx - \frac{\epsilon}{N} \sum_{i,a} \ln (e^{\frac{|\vect{H}_{ia}|}{\epsilon}}) \approx - \frac{1}{N} \sum_{i,a} |\vect{H}_{ia}|. \label{eq:EntropyField}
\end{align}
When temperature becomes finite, the entropic contribution is therefore related to the average magnitude of the local fields.

As shown in Eq. \eqref{eq:LocalField}, there are two contributions to this local field: the local field resulting from the exchange between moments, and the local field resulting from the Zeeman coupling to the applied field ${\bm B}$. When ${\bm B} = 0$, only the first contribution is active. For the exchange parameters used here, the exchange couplings create local fields that move the spins into their Palmer-Chalker states, hence making these the ground state configurations. When $B$ is turned on, we must consider the second contribution, given by Eq.~\eqref{eq:ZeemanField}.  Assuming a pure easy-plane anisotropy on each sublattice, there are four $g$-tensors to consider \cite{MPCPyrochlore}:
\begin{align}
    \overleftrightarrow{g_0} &= \frac{g_\perp}{3} \begin{pmatrix}2 && -1 && -1 \\ -1 && 2 && -1 \\ -1 && -1 && 2 \end{pmatrix}, 
    \overleftrightarrow{g_1} = \frac{g_\perp}{3} \begin{pmatrix}2 && 1 && 1 \\ 1 && 2 && -1 \\ 1 && -1 && 2 \end{pmatrix}, \nonumber \\
    \overleftrightarrow{g_2} &= \frac{g_\perp}{3} \begin{pmatrix}2 && 1 && -1 \\ 1 && 2 && 1 \\ -1 && 1 && 2 \end{pmatrix},
    \overleftrightarrow{g_3} = \frac{g_\perp}{3} \begin{pmatrix}2 && -1 && 1 \\ -1 && 2 && 1 \\ 1 && 1 && 2 \end{pmatrix}.  \nonumber
\end{align}
For a [110] field, a remarkable coincidence occurs: the Zeeman contribution $\mu_{\textrm B} g_{ia}^{\mu \nu} B^{\nu}$ for two sublattices (namely, sublattices 1 and 2) lie within the $xy$ plane, exactly parallel to the spin configuration of the $\langle xy \rangle$ and $\langle \overline{xy} \rangle$ Palmer-Chalker states in Table \ref{tbl:PCStates}. As such, for low fields in the [110] direction, these spins \textit{do not} cant out of their original positions; the local fields from the exchange and Zeeman contributions add in parallel. Returning to Eq. \eqref{eq:EntropyField}, this lack of canting is the origin of the slight entropy difference between the Palmer-Chalker states that lie in the $xy$-plane (where the two contributions add in parallel) and out of the $xy$-plane (where the two contributions are not parallel). A similar discrepancy should be expected at $T=0$: $\vect{m}_{ia} = \frac{\vect{H}_{ia}}{|\vect{H}_{ia}|}$ at $T=0$, so this lack of canting will discriminate between the $xy$-planar and non-$xy$-planar states. This effect is therefore the origin of the difference between the $T=0$ and $T=0^+$ phases. It should be noted that this is \textit{not} an order by disorder effect. The $\mathbb{Z}_2$ symmetry found at low fields and at $T=0^+$ is not a subset of the original $\mathbb{Z}_4$ symmetry found at low fields and at $T=0$. Rather, the $\mathbb{Z}_2$ symmetry corresponds to the two $xy$-planar Palmer-Chalker states, whereas the $\mathbb{Z}_4$ symmetry corresponds to the \textit{other} four Palmer-Chalker states.

\clearpage
\section{Monte Carlo simulations at low field in the [110] direction}
\label{SM-Analysis110MCsmallfield}

\begin{figure}[ht]
\centering\includegraphics[width=17cm]{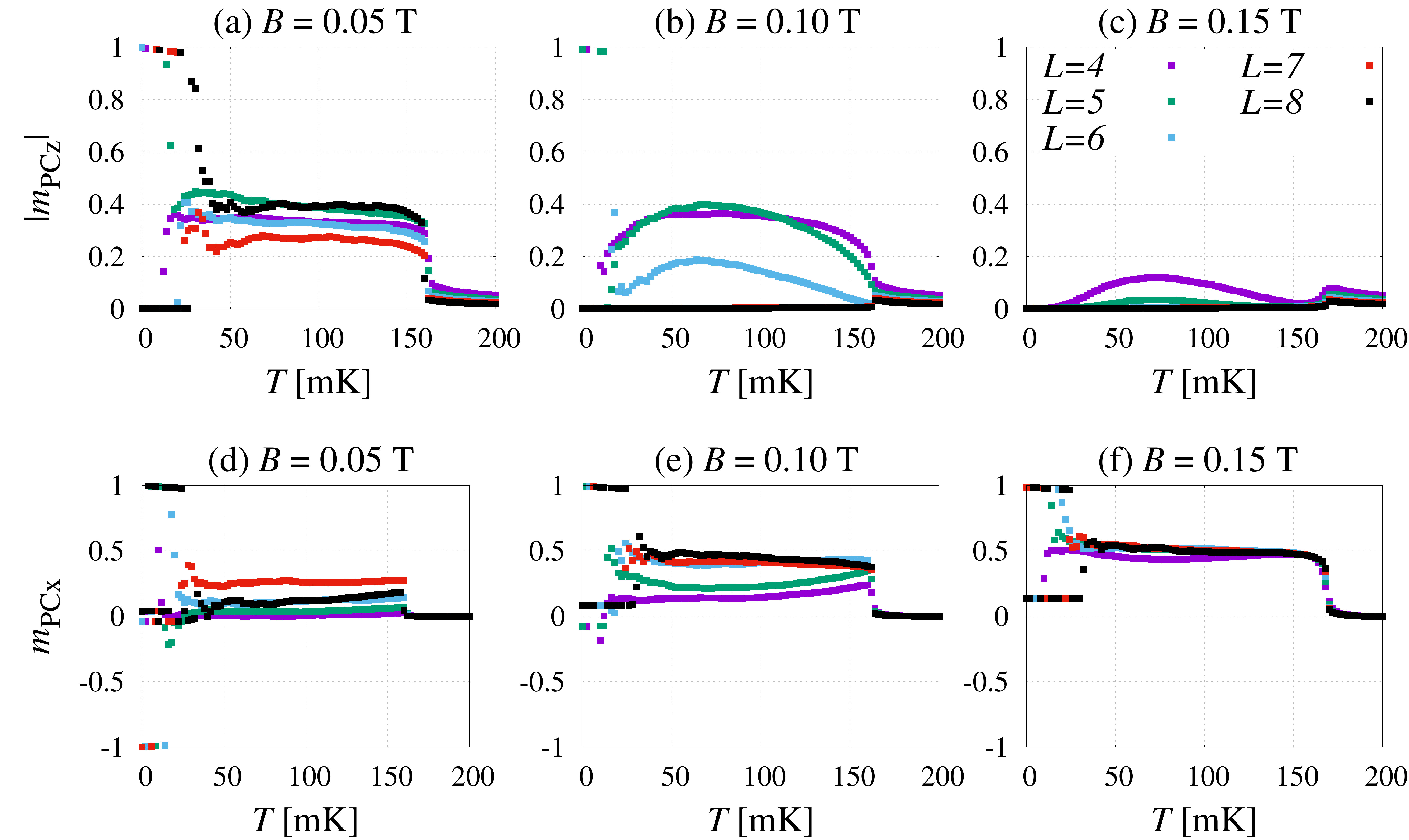}
\caption{
At low field in the $[110]$ direction, there are four FEPC ground states $\{\langle xz \rangle_h, \langle yz \rangle_h, \langle \overline{xz} \rangle_h, \langle \overline{yz} \rangle_h\}$ and two FEPC excited states $\{\langle xy \rangle_h, \langle \overline{xy} \rangle_h\}$ [see Fig.~\ref{fig:VMFTResults}]. The very small energy gap between them means that it is not obvious which order is stabilised at finite temperature.\\
The easiest way to differentiate between the ground and excited states in Monte Carlo simulations is via the $z$-component of the Palmer-Chalker order parameter, $m_{\rm PCz}$ [see (a,b,c) panels]. $m_{\rm PCz}$ is finite for the two excited states, while it is zero for the four ground states. Here we show that as the system size is increased, $m_{\rm PCz}$ vanishes for $B=$ 0.1 and 0.15 T, which means the FEPC excited states are not stable for $B\gtrsim 0.1$ T. For $B=$ 0.05 T on the other hand, simulations are difficult to thermalize and the evolution with system size is not monotonic. Hence, we cannot rule out the presence of the FEPC excited states at finite temperature, as predicted by mean field theory in the previous section. That being said, this possible co-existence would only arise at very low field, $B\lesssim 0.1$ T, and thus not affect the mechanism for reentrance.\\
As for the four FEPC ground states, they can be divided into two pairs. For $\langle \overline{xz} \rangle_h$ and $\langle \overline{yz} \rangle_h$, the $x$-component of the Palmer-Chalker order parameter is positive, $m_{\rm PCx}>0$. For $\langle xz \rangle_h$ and $\langle yz \rangle_h$, $m_{\rm PCx}<0$. As shown in panels (d,e,f), $m_{\rm PCx}$ always gets more and more positive for large system sizes, which means that the Z$_{2}$ states at low $[110]$ field are $\langle \overline{xz} \rangle_h$ and $\langle \overline{yz} \rangle_h$. At very low temperatures, $T< 40$ mK, the data split into two groups because of the broken ergodicity in simulations between $\langle \overline{xz} \rangle_h$ and $\langle \overline{yz} \rangle_h$.\\
All of these simulations were done for $t_{m}=5. 10^{7}$ Monte-Carlo steps.
}
\label{fig:110PCSM}
\end{figure}
\clearpage

\section{Monte Carlo simulations in the [111] direction}
\label{SM-Analysis111MC}

\begin{figure}[ht]
\centering\includegraphics[width=17cm]{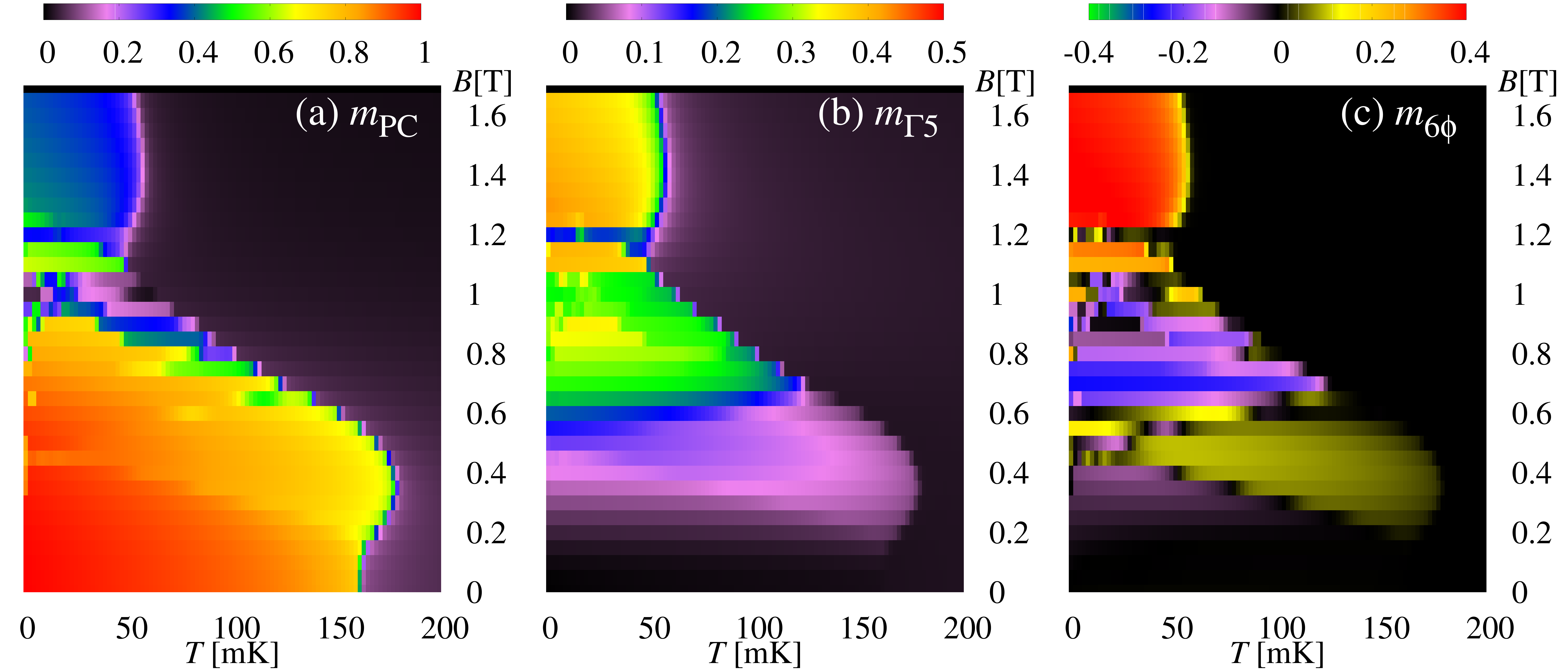}
\caption{
Here we reproduce the phase diagrams in a $[111]$ field as signalled  via the (a) Palmer-Chalker and (b) $\Gamma_{5}$ order parameters, as well as the quantity (c) $m_{6\phi}$ (defined below) to differentiate between $\psi_{2}$ and $\psi_{3}$ states. The contour of the phase diagram is clearly visible in panels (a) and (b). However, as explained in the main text, in order to understand the reentrance at $B\sim 0.5$ T, we need to consider the evolution of the angle $\phi$. The $\psi_{2}$ ($\psi_{3}$) states are characterized by $\phi \equiv n \pi/3\;(+\pi/6)$ for $n=0,...,5$. Hence, the quantity $m_{6\phi}\equiv m_{\Gamma 5} \cos\left(6\phi\right)$ is equal to +1 (-1) for $\psi_{2}$ ($\psi_{3}$) states. The yellow region at finite temperature for $B\sim 0.5$ T thus indicates the dominance of $\psi_{2}$ order in this region, as discussed in the main text.\\
All of these simulations were done for $L=6$ and $t_{m}=5. 10^{7}$ Monte-Carlo steps.
}
\label{fig:111SM}
\end{figure}
